\documentclass[
twocolumn,
preprintnumbers,
 amsmath,amssymb,
 aps,
 superscriptaddress
]{revtex4}
\usepackage{xcolor}
\usepackage{graphicx}
\usepackage{dcolumn}
\usepackage{bm}
\usepackage{comment}
\usepackage{algorithm}
\usepackage{algorithmic}

\begin{document}

\title{Estimating time-dependent entropy production from non-equilibrium trajectories}

\author{Shun Otsubo}
\affiliation{
 Department of Applied Physics, The University of Tokyo, 7-3-1 Hongo, Bunkyo-ku, Tokyo 113-8656, Japan
}

\author{Sreekanth K Manikandan}
\affiliation{Department of Physics, Stockholm University, SE-10691 Stockholm, Sweden}
\affiliation{NORDITA, Royal Institute of Technology and Stockholm University,
Roslagstullsbacken 23, SE-10691 Stockholm, Sweden}

\author{Takahiro Sagawa}
\affiliation{
 Department of Applied Physics, The University of Tokyo, 7-3-1 Hongo, Bunkyo-ku, Tokyo 113-8656, Japan
}
\affiliation{
Quantum-Phase Electronics Center (QPEC), The University of Tokyo, 7-3-1 Hongo, Bunkyo-ku, Tokyo 113-8656, Japan
}

\author{Supriya Krishnamurthy}%
\affiliation{Department of Physics, Stockholm University, SE-10691 Stockholm, Sweden}%

\date{\today}

\begin{abstract}
The rate of entropy production provides a useful quantitative measure of a non-equilibrium system and estimating it directly from time-series data from experiments is highly desirable.
Several approaches have been considered for stationary dynamics, some of which are based on a variational characterization of the entropy production rate. However, the issue of obtaining it in the case of non-stationary dynamics remains largely unexplored.
Here, we solve this open problem by demonstrating that the variational approaches can be generalized to give the exact value of the entropy production rate even for non-stationary dynamics.
On the basis of this result, we develop an efficient algorithm that estimates the entropy production rate continuously in time by using machine learning techniques, 
and validate our numerical estimates using analytically tractable Langevin models \textcolor{black}{in experimentally relevant parameter regimes}. Our method is of great practical significance since all it requires is time-series data for the system of interest without requiring prior knowledge of the system parameters.
\end{abstract}

\maketitle\begin{figure*}[t]
\includegraphics[width = 0.95\linewidth]{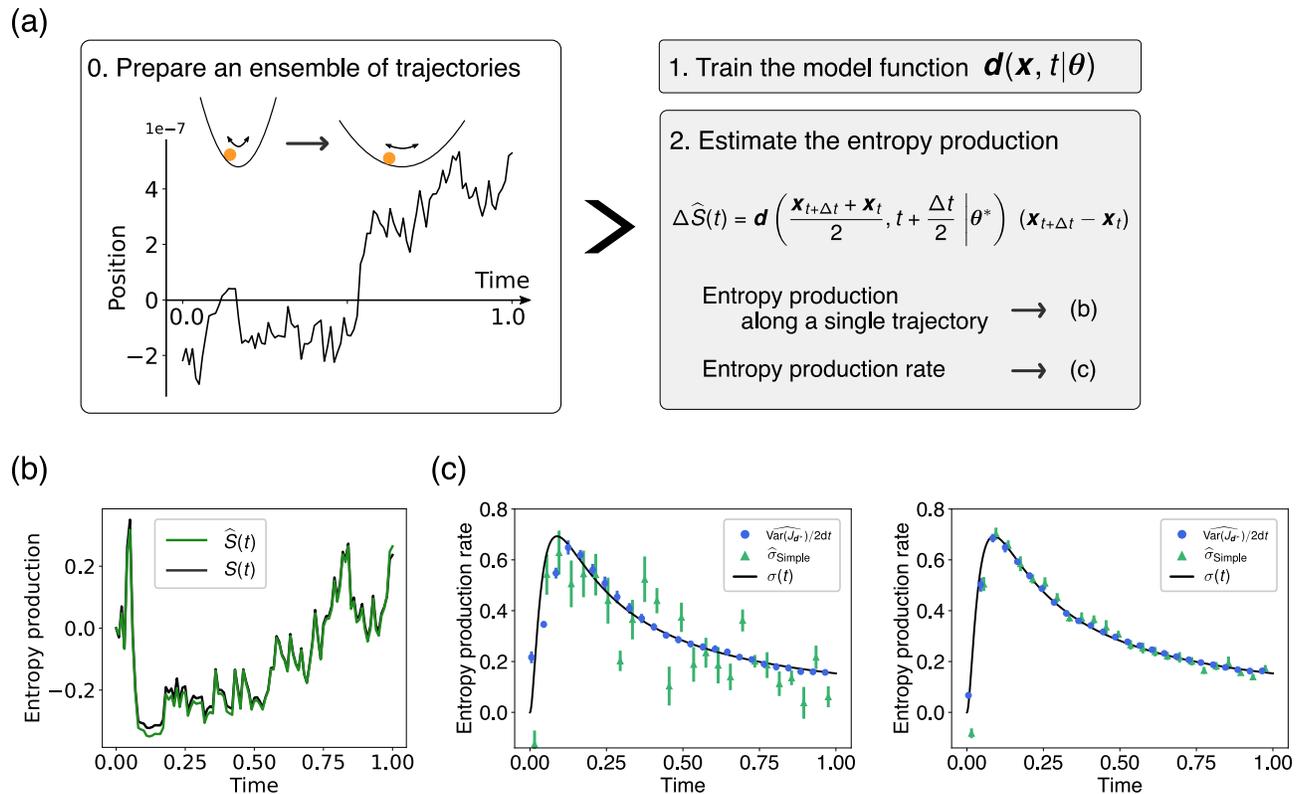}
\caption{Estimating the entropy production along non-stationary trajectories: (a) Schematic of our inference scheme. First, we prepare an ensemble of trajectories generated by an experiment or the equation of interest. Then, we optimize the model function of the coefficient field ${\bm d}({\bm x}, t | {\bm\theta})$ to get an estimate of the thermodynamic force ${\bm F}({\bm x}, t)$, and use it to estimate the entropy production along each trajectory or the (ensemble-averaged) entropy production rate. In the left box, we show an example of a trajectory generated by the breathing parabola model.
(b) Estimated entropy production along a single trajectory. The thin green line is the estimated entropy production, and the thick black line is the true entropy production calculated analytically. The estimation is conducted for the trajectory depicted in panel (a) \textcolor{black}{after training the model function using $10^5$ trajectories.}
(c) Estimated entropy production rate. \textcolor{black}{The blue circles and green triangles are the estimated values using $10^4$ (left) or $10^5$ trajectories (right)}, and the black line is the true entropy production rate. \textcolor{black}{We use the variance-based estimator (Eq.~(\ref{eq: variance-based})) for the blue circles and the simple dual representation (Eq.~(\ref{eq: simple})) for the green triangles after training the model function with the simple dual representation in both the cases. The variance-based estimator reduces the statistical error significantly.}
The mean of 10 independent trials and its standard deviation are plotted for the estimated values. For (a)-(c), trajectories are generated by the breathing parabola model (Eq.\ \eqref{eq:BreathingP}) with \textcolor{black}{$\alpha = 11.53$, $D=1.6713\times 10^{-13}$, $\Delta t = 10^{-2}$, and $\tau_{\rm obs} = 1$.}}
\label{fig: main1}
\end{figure*}

\section*{Introduction}
The entropy production rate is an important quantitative measure
of a non-equilibrium process and knowing its value is indicative of useful information about the system such as heat dissipated \cite{harada2005equality,toyabe2007experimental},
efficiency (if the non-equilibrium system in question is an engine \cite{Verley:2014ute,Verley:2014uce,Manikandan:EF})
as well as free energy differences \cite{JarzynskiPRL,Crooks1999} (if the non-equilibrium process interpolates between two equilibrium states).
In particular, the entropy production rate often characterizes the energy consumption of non-equilibrium systems \cite{Seara2018}.
It also provides useful information for  systems with hidden degrees of freedom \cite{Esposito2012, Kawaguchi2013}, or interacting subsystems where information theoretic quantities play a key role \cite{Sagawa2012, Ito2013, Horowitz2014, Parrondo2015}.\\ \indent
The entropy production rate can be directly obtained from the system's phase-space trajectory if the underlying dynamical equations of the system are known \cite{sekimoto1997kinetic,sekimoto1998langevin,Seifert:2005epa,Seifert:2012stf}.  
 This is not the case however for the vast majority of systems, such as biological systems \cite{battle2016broken,fang2019nonequilibrium,matsumoto2018role}, and
 consequently, there has been a lot of interest in developing new methods for estimating the entropy production rate directly from trajectory data \cite{roldan2010estimating, Seifert:inf, noninvasive, Martinez2019, Gingrich:qua,manikandan2019inferring,Shun:eem, van:epe, kim2020learning, Frishman2020, Gnesotto2020, Kappler2020}. Some of these techniques involve the estimation of the probability distribution and currents over the phase-space \cite{roldan2010estimating,Gingrich:qua}, which requires huge amounts of data. Some other techniques are invasive and require perturbing the system  \cite{harada2005equality,toyabe2007experimental}, which may not always be easy to implement.  \\ \indent
An alternative strategy is to set lower bounds on the entropy production rate \cite{Kawai2007, Blythe2008, Vaikuntanathan2009, Muy2013,PhysRevLett.114.158101} by measuring experimentally accessible quantities. 
One class of these bounds, for example those based on the thermodynamic uncertainty relation (TUR) \cite{ PhysRevLett.114.158101, Gingrich2016, Horowitz2017, Gingrich2017, horowitz2019thermodynamic}, have been further developed into variational {\em inference} schemes which translate the task of identifying entropy production to an optimization problem over the space of a single projected fluctuating current in the system \cite{Gingrich:qua,manikandan2019inferring,Shun:eem,van:epe}. Recently a similar variational scheme using neural networks was also proposed \cite{kim2020learning}. As compared to other trajectory-based entropy estimation methods, these inference schemes do not involve the estimation of any kind of empirical distributions over the phase-space, and are hence known to work better in higher dimensional systems \cite{Gingrich:qua}.
In addition, it is proven that such an optimization problem gives the exact value of the entropy production rate in a stationary state if short-time currents are used \cite{manikandan2019inferring,Shun:eem,van:epe,kim2020learning}.
The short-time TUR has also been experimentally tested in colloidal particle systems recently \cite{Manikandan2021}.
However, whether these existing schemes work well for non-stationary states has not been explored as yet.\\ \indent
Non-stationary dynamics ubiquitously appear in biological phenomena such as in adaptive responses to environmental change \cite{Lan2012} and spontaneous oscillations \cite{Zambrano2016}, all of which are inevitably accompanied by energy dissipation.
However, for a non-stationary system, it has only been possible to place bounds on the time-dependent entropy produced during a finite time-interval under specific \cite{Koyuk:op,liu2019thermodynamic} or more general \cite{koyuk2020thermodynamic}  conditions.
In addition, there is no guarantee that these bounds can be saturated by 
any quantity related to the entropy production of the system.
Hence there is no established scheme that has been proven to work for obtaining the exact entropy production rate under time-dependent conditions.
\\ \indent
We address this outstanding problem by proposing a class of variational inference schemes which can give not only the exact value of the time-dependent entropy production rate under non-stationary
conditions but even entropy production along single realizations.
These schemes, which can be directly implemented on time-series data obtained from experiments,
involve maximization over an objective function which consists of a single projected current determined from the data.
We demonstrate that this objective function can either be of the form dictated by the recently proposed short-time thermodynamic uncertainty relation \cite{manikandan2019inferring,Shun:eem,van:epe} or the form recently suggested in \cite{kim2020learning}, or a variation of these. The collection of these schemes work for both diffusive systems described by overdamped Langevin equations as well as finite-state-space systems described by master equations and work for both transient as well as stationary states.  \\ \indent
We implement these variational schemes by means of an efficient algorithm that estimates the entropy production continuously in time by modeling the time-dependent projection coefficients with a feedforward neural network and by carrying out gradient ascent using machine learning techniques.
This algorithm can in principle be directly used on real experimental data.
Here, however, as a proof of concept, we consider time-series data generated by two models; one of a colloidal particle in a time-varying trap and the other of a biological model that describes biochemical reactions affected by a time-dependent input signal, for both of which we can obtain exact solutions for the time-dependent entropy production rate as well as the  entropy production along single trajectories.
We then demonstrate that our proposed scheme indeed works by comparing the numerical implementation  to our theoretical predictions (see Fig.\ref{fig: main1}).\\ \indent

\section*{Results}
\subsection*{Short-time variational representations of the entropy production rate}

The central results we obtain, summarized in Fig.~\ref{fig: main1}, are applicable to experimental data from any non-equilibrium system,
at least in principle, described by an overdamped Langevin equation or a Markov jump process even without knowing any details of the equations involved. 
Here we use the model of a generic overdamped Langevin dynamics in $d$-dimensions in order to introduce the notations. We consider an equation of the form:
\begin{align}
    \dot{{\bm x}}(t)={\bm A}({\bm x}(t),t)+{\bm B}({\bm x}(t), t)\cdot {\bm \eta}(t),\label{eq: Langevin}
\end{align}
where ${\bm A}({\bm x}, t)$ is the drift vector, and ${\bm B}({\bm x}, t)$ is a $d \times d$ matrix, and ${\bm \eta}(t)$ represents a Gaussian white noise satisfying $\langle \eta_i(t) \eta_j(t')\rangle = \delta_{ij}\delta(t-t')$. Note that we adopt the Ito-convention for the multiplicative noise.
The corresponding Fokker-Planck equation satisfied by the probability density $p({\bm x},t)$ reads
\begin{align}
\label{eq:generic}
    \partial_t p({\bm x}, t) &= -{\bm \nabla}\;{\bm j}({\bm x}, t),\\
    j_i({\bm x}, t) &= A_i({\bm x}, t) p({\bm x}, t) - \sum_j\nabla_j\left[D_{ij}({\bm x}, t)p({\bm x}, t) \right],\label{eq:prob_cur}
\end{align}
where $\bm D$ is the diffusion matrix defined by
\begin{align}
    {\bm D}({\bm x}, t)=\frac{1}{2}{\bm B}({\bm x}, t){\bm B}({\bm x}, t)^T
\end{align}
and ${\bm j}({\bm x},t)$ is the probability current. Equations of the form Eq.\ \eqref{eq:generic} can, for example, be used to describe the motion of colloidal particles in optical traps \cite{bpexpt,opt,spexpt,sphs}. In some of these cases, the Fokker-Planck equation can also be solved exactly to obtain the instantaneous probability density $p({\bm x},t)$.\\ \indent 
Whenever ${\bm j}({\bm x},t)\neq 0$, the system is out of equilibrium. How far the system is from equilibrium can be quantified using the average rate of the entropy production at a given instant $\sigma(t)$, which can be formally obtained as the integral \cite{Spinney2012} 
\begin{align}
\label{eq:sigmatime}
    \sigma(t)=\int {\rm d}{\bm x}\; {\bm F}({\bm x},t)\;{\bm j}({\bm x},t),
\end{align}
 where ${\bm F}({\bm x},t)$ is the thermodynamic force defined as
 \begin{align}
 \label{eq:Ffield}
 {\bm F}({\bm x},t) &= \frac{{\bm j}^T({\bm x},t){\bm D}({\bm x}, t)^{-1}}{p({\bm x},t)}.
 \end{align}
Note that the Boltzmann's constant is set to unity $k_B = 1$ throughout this paper. Further, the entropy production along a stochastic trajectory denoted as $S[{\bm x}(\cdot),t]$ can be obtained as the integral of the single-step entropy production
 \begin{align}
    {\rm d}S =  {\bm F}\left({\bm x}(t), t\right)\circ {\rm d}{\bm x}(t) \label{eq: single_ep},
 \end{align}
where $\circ$ denotes the Stratonovich product. This quantity is related to the average entropy production rate as $\sigma(t) = \langle {\rm d}S(t)/{\rm d}t\rangle$, where $\langle\cdots\rangle$ denotes the ensemble average.
Similar expressions can be obtained for any Markov jump processes if the underlying dynamical equations are specified \cite{Seifert:2005epa}.\\ \indent
In the following we discuss two variational representations that can estimate $\sigma(t)$, ${\bm F}({\bm x},t)$ and $S[{\bm x}(\cdot),t]$ in non-stationary systems, without requiring the prior knowledge of the dynamical equation. We also construct a third simpler variant, and comment on the pros and cons of these different representations for inference.\\ \indent

\begin{table}
\begin{center}
\begin{tabular}{ c | c c c c} \hline \hline
 Rep.						& Markov jump & Langevin & Optimal field & Tightness \\ \hline
$\sigma_{\rm NEEP}$	& Yes	 	& Yes			      & ${\bm d}^*({\bm x}) = {\bm F}({\bm x}, t)$   & Loose\\
$\sigma_{\rm Simple}$	& No			& Yes			      & ${\bm d}^*({\bm x}) = {\bm F}({\bm x}, t)$   & Loose\\		   
$\sigma_{\rm TUR}$		& No			& Yes			      & ${\bm d}^*({\bm x}) \propto{\bm F}({\bm x}, t)$ & Tight\\ \hline \hline
\end{tabular}
\caption{Summary of the comparison among the variational representations $\sigma_{\rm NEEP}$, $\sigma_{\rm Simple}$ and $\sigma_{\rm TUR}$.}
\label{table: summary}
\end{center}
\end{table}

\noindent
\textbf{TUR representation.}
The first method is based on the TUR \cite{Gingrich:qua, PhysRevLett.114.158101, Gingrich2016, Horowitz2017, Gingrich2017, horowitz2019thermodynamic}, which provides a lower bound for the entropy production rate in terms of the first two cumulants of non-equilibrium current fluctuations directly measured from the trajectory. It was shown recently that the TUR provides not only a bound, but even an exact estimate of the entropy production rate for stationary overdamped Langevin dynamics by taking the short-time limit of the current \cite{manikandan2019inferring,Shun:eem,van:epe}. Crucially, the proof in Ref.~\cite{Shun:eem} is also valid for non-stationary dynamics.\\ \indent
This gives a variational representation of the entropy production rate, given by the estimator
\begin{eqnarray}
\label{eq:TURinf}
\sigma_{\rm TUR}(t) := \frac{1}{{\rm d}t}\max_{{\bm d}}\frac{2\left<J_{\bm d}\right>^2}{{\rm Var}(J_{\bm d})},
\end{eqnarray}
where $J_{\bm d}$ is the (single-step) generalized current given by $J_{\bm d}:= {\bm d}({\bm x}(t))\circ {\rm d}{\bm x}(t)$ defined with some coefficient field ${\bm d}({\bm x})$. The expectation and the variance are taken with respect to the joint probability density $p({\bm x}(t), {\bm x}(t+{\rm d}t))$. In the ideal short-time limit ${\rm d}t\to 0$, the estimator gives the exact value, i.e., $\sigma_{\rm TUR}(t) = \sigma(t)$ holds \cite{Shun:eem}.
The optimal current that maximizes the objective function is proportional  to the entropy production along a trajectory, $J^*_{\bm d} = c {\rm d}S$, and the corresponding coefficient field is ${\bm d}^*({\bm x}) = c{\bm F}({\bm x}, t)$, where the constant factor $c$ can be removed by calculating $2\left<J_{\bm d}\right>/{\rm Var}(J_{\bm d}) = 1/c$.\\ \indent

\noindent
\textbf{NEEP representation.}
The second variational scheme is the Neural Estimator for Entropy Production (NEEP) proposed in Ref.~\cite{kim2020learning}. In this study, we define the estimator $\sigma_{\rm NEEP}$ in the form of a variational representation of the entropy production rate as
\begin{eqnarray}
\sigma_{\rm NEEP}(t) := \frac{1}{{\rm d}t}\max_{{\bm d}}\left< J_{\bm d} - e^{-J_{\bm d}} + 1\right>,\label{eq: NEEP}
\label{eq:NEEP}
\end{eqnarray}
where the optimal current is the entropy production itself, $J_{\bm d}^* = {\rm d}S$, and the corresponding coefficient field is ${\bm d}^*({\bm x}) = {\bm F}({\bm x}, t)$.
Again, in the ideal short-time limit, $\sigma_{\rm NEEP}(t) = \sigma(t)$ holds. Eq.~(\ref{eq:NEEP}) is a slight modification of the variational formula obtained in Ref.~\cite{kim2020learning}; we have added the third term so that the maximized expression itself gives the entropy production rate.
Although it was derived for stationary states there, it can be shown that such an assumption is not necessary in the short-time limit. We provide a proof of our formula using a dual representation of the Kullback-Leibler divergence \cite{Keziou2003,Nguyen2010,Belghazi2018} in the Supplementary Information.\\ \indent
In contrast to the TUR representation, NEEP requires the convergence of  exponential averages of current fluctuations, but it provides an exact estimate of the entropy production rate not only for  diffusive Langevin dynamics but also for any Markov jump process. Since there are some differences in the estimation procedure for these cases \cite{Shun:eem,kim2020learning}, we focus on  Langevin dynamics in the following, while its use in Markov jump processes is discussed in the Supplementary Information. \\ \indent

\noindent
\textbf{Simple dual representation.}
For Langevin dynamics, we also derive a new representation, named the simple dual representation $\sigma_{\rm Simple}$ by simplifying $\left<e^{-J_{\bm d}}\right>$ in the NEEP estimator as
\begin{eqnarray}
\sigma_{\rm Simple}(t) := \frac{1}{{\rm d}t}\max_{{\bm d}}\left[2\left<J_{\bm d}\right> - \frac{{\rm Var}(J_{\bm d})}{2}\right].\label{eq: simple}
\end{eqnarray}
Here, the expansion of $\left<e^{-J_{\bm d}}\right>$ in terms of the first two moments is exact only for Langevin dynamics and hence this representation cannot be used for Markov jump processes \textcolor{black}{to obtain $\sigma$ (however, as shown in \cite{MacIeszczak2018}, the equivalence of the TUR objective function to the objective function in the above representation continues to hold in the long-time limit).} The tightness of the simple dual and TUR bounds can be compared as follows: In Langevin dynamics, for any fixed choice of $J_{\bm d}$,
\begin{eqnarray}
\sigma {\rm d}t\geq\frac{2\left<J_{\bm d}\right>^2}{{\rm Var}(J_{\bm d})} \geq 2\left<J_{\bm d}\right> - \frac{{\rm Var}(J_{\bm d})}{2},\label{eq: comparison}
\end{eqnarray}
where we used the inequality $\frac{2a^2}{b}\geq 2a - \frac{b}{2}$ for any $a$ and $b>0$.
Since a tighter bound is advantageous for the estimation \cite{Ruderman2012, Belghazi2018}, $\sigma_{\rm TUR}$ would be more effective for estimating the entropy production rate for the Langevin case.\\ \indent
On the other hand, $\sigma_{\rm NEEP}$ and $\sigma_{\rm Simple}$ have an advantage over $\sigma_{\rm TUR}$ in estimating the thermodynamic force ${\bm F}({\bm x}, t)$, since the optimal coefficient field is the thermodynamic force itself for these estimators. In contrast, $\sigma_{\rm TUR}$ needs to cancel the constant factor $c$ by calculating $2\left<J_{\bm d}\right>/{\rm Var}(J_{\bm d}) = 1/c$, which can increase the statistical error due to the fluctuations of the single-step current (see the Supplementary Information for further discussions and numerical results). 
In the next section, we propose a continuous-time inference scheme that estimates in one shot,
the time-dependent thermodynamic force for the entire time range of interest. This results in an accurate estimate with less error than the fluctuations of the single-step current.
$\sigma_{\rm NEEP}$ and $\sigma_{\rm Simple}$ are more effective for this purpose, since the correction of the constant factor $c$, whose expression is based on the single-step current, negates the benefit of the continuous-time inference for $\sigma_{\rm TUR}$. 
In Table 1, we provide a summary of the three variational representations.\\ \indent
We note that the variational representations are exact only when all the degrees of freedom are observed; otherwise they give a lower bound on the entropy production rate. This can be understood as an additional constraint on the optimization space. For example, when the $i$-th variable is not observed, it is equivalent to dropping $x_i$ from the argument of ${\bm d}({\bm x})$ and setting $d_i = 0$.
We also note that the variational representations are exact to  order ${\rm d}t$; in practice, we use a short but finite ${\rm d}t$.  The only variational representation which can give the exact value with any finite ${\rm d}t$ is $\sigma_{\rm NEEP}$, under the condition that the dynamics is stationary \cite{kim2020learning}.\\ \indent

\subsection*{An algorithm for \textcolor{black}{non-stationary} inference}
The central idea of our inference scheme is depicted in Fig.~\ref{fig: main1}(a).
Equations (\ref{eq:TURinf}), (\ref{eq: NEEP}) and (\ref{eq: simple}) all give the exact value of $\sigma (t)$ in principle in the Langevin case, but here we elaborate on how we implement them in practice. 
We first prepare an ensemble of finite-length trajectories, which are sampled from a non-equilibrium and non-stationary dynamics with $\Delta t$ as the sampling interval:
\begin{eqnarray}
\Gamma_i = \left\{{\bm x}^{(i)}_{0}, {\bm x}^{(i)}_{\Delta t}, ..., {\bm x}^{(i)}_{\tau_{\rm obs}} (= {\bm x}^{(i)}_{M\Delta t}) \right\}~(i = 1, ..., N) .
\end{eqnarray}
Here $i$ represents the index of trajectories, $N$ is the number of trajectories, and $M$ is the number of transitions. The subscript $(i)$ will be often omitted for simplicity. Then, we estimate the entropy production rate $\sigma(t)$ using the ensemble of single transitions $\lbrace{{\bm x}_t, {\bm x}_{t+\Delta t} \rbrace}_i$ at time $t$. $\sigma(t)$ is obtained by finding the optimal current that maximizes the objective function which is itself estimated using the data. Hereafter, we use the hat symbol for quantities estimated from the data: for example, $\widehat{\sigma}_{\rm Simple}(t)$ is the estimated objective function of the simple dual representation. We also use the notation $\widehat{\sigma}(t)$ when the explanation is not dependent on the particular choice of the representation.
The time interval for estimating $\widehat{\sigma}(t)$ is set to be equal to the sampling interval $\Delta t$ for simplicity, but they can be different in practice, i.e., transitions $\{{\bm x}_t, {\bm x}_{t + n\Delta t}\}$ with some integer $n\geq 1$ can be used to estimate $\widehat{\sigma}(t)$ for example.
\\ \indent
Concretely, we can model the coefficient field with a parametric function ${\bm d}({\bm x} | {\bm \theta})$ and conduct the gradient ascent for the parameters $\bm\theta$. 
As will be explained, we use a feedforward neural network for the model function, where ${\bm\theta}$ represents, for example, weights and biases associated with nodes in the neural network. In this study, we further optimize the coefficient field continuously in time, i.e., optimize a model function ${\bm d}({\bm x}, t | {\bm \theta})$ which includes time $t$ as an argument. 
The objective function to maximize is then given by
{\color{black}\begin{eqnarray}
f({\bm \theta}) := \frac{1}{M}\sum_{j=0}^{M-1} \widehat{\sigma}(j\Delta t).\label{eq: objective_func}
\end{eqnarray}}
The optimal model function ${\bm d}({\bm x}, t|{\bm \theta}^*)$ that maximizes the objective function is expected to approximate well the thermodynamic force ${\bm F}({\bm x}, t)$ (or $c(t){\bm F}({\bm x}, t)$ if $\sigma_{\rm TUR}$ is used) at least at $j\Delta t ~(j = 0, 1, ...)$, 
and even at interpolating  times if $\Delta t$ is sufficiently small.
Here, ${\bm \theta}^*$ denotes the set of optimal parameters obtained by the gradient ascent, and we often use ${\bm d}^*$ to denote the optimal model function ${\bm d}({\bm x}, t|{\bm \theta}^*)$ hereafter.\\ \indent 
This continuous-time inference scheme is a generalization of the instantaneous-time inference scheme. \textcolor{black}{Instead of optimizing a time-independent model function ${\bm d}({\bm x}|{\bm\theta})$ in terms of $\widehat{\sigma}(j\Delta t)$ with a fixed index $j$, the continuous-time scheme needs to perform only one optimization of the sum Eq.~(\ref{eq: objective_func}). This makes it
much more data efficient in utilizing the synergy between ensembles of single transitions at different times.
This also ensures that we can get the smooth change of the thermodynamic force, interpolating discrete-time transition data.}\\

\noindent
\textbf{Variance-based estimator.}
\textcolor{black}{
In principle, all the three variational representations work as an estimator of the entropy production rate as well. However, as we detail in the Supplementary Information, once we have obtained an estimate of the thermodynamic force ${\bm d^*}\simeq{\bm F}$ (taking into account the correction term for $\widehat{\sigma}_{\rm TUR}$) by training the model function, it is possible to use a variance-based estimator of the entropy production rate,
\begin{eqnarray}
\frac{1}{2{\rm d}t}{\rm Var}(J_{\bm d^*})\simeq\frac{1}{2{\rm d}t}{\rm Var}(J_{\bm F}) = \sigma(t),
\label{eq: variance-based}
\end{eqnarray}
which can considerably reduce the statistical error. This is due to the fact that $\widehat{\left<J_{\bm d}\right>}$ fluctuates around $\left<J_{\bm d}\right>$ more than $\widehat{{\rm Var}(J_{\bm d})}$ does around ${\rm Var}(J_{\bm d})$ for any choice of ${\bm d}$, for small ${\rm d}t $ (see the Supplementary Information for the derivation).
The above  advantage in using the variance as an estimator, instead of the mean, would normally be masked by noise in the estimation of ${\bm d^*}$. However, if the coefficient field is trained by $\widehat{\sigma}_{\rm Simple}$ or $\widehat{\sigma}_{\rm NEEP}$ with the continuous-time inference scheme, remarkably, ${\bm d^*}$ is obtained with an accuracy beyond the statistical error of $\widehat{\left<J_{\bm d}\right>}$ since it takes the extra constraint of time-continuity into account. This results in the error of $\widehat{{\rm Var}(J_{\bm d^*})}$ being smaller than that of $\widehat{\left<J_{\bm d^*}\right>}$, because of the difference in how the leading-order terms of their statistical fluctuation scale with ${\rm d}t$.
We note that $\widehat{\sigma}_{\rm TUR}$ is not appropriate for this purpose, since in this case, ${\bm d}^*$ should be multiplied by $2\widehat{\left<J_{\bm d}\right>}/\widehat{{\rm Var}(J_{\bm d})}$ to obtain an estimate of the thermodynamic force, which increases the statistical error to the same level as $\widehat{\left<J_{\bm d}\right>}$. \\ \indent
In numerical experiments, we mainly use $\widehat{\sigma}_{\rm Simple}$ for training the coefficient field to demonstrate the validity of this new representation, and use the variance-based estimator for estimating the entropy production rate.}\\ \indent

We adopt the data splitting scheme \cite{Shun:eem, kim2020learning} for training the model function to avoid the underfitting and overfitting of the model function to trajectory data.
Concretely, we use only half the number of trajectories for training the model function, while we use the other half for evaluating the model function and estimating the entropy production.
In this scheme, the value of the objective function calculated with the latter half (we call it test value) quantifies the generalization capability of the trained model function.
Thus, we can compare two model functions, and expect that the model function with the higher test value gives the better estimate.
We denote the optimal parameters that maximize the test value during the gradient ascent as ${\bm \theta}^*$.
Hyperparameter values are obtained similarly.
Further details, including a pseudo code, are provided in the Methods section.

\subsection*{Numerical results}
\begin{figure*}[t]
\includegraphics[width = 0.85\linewidth]{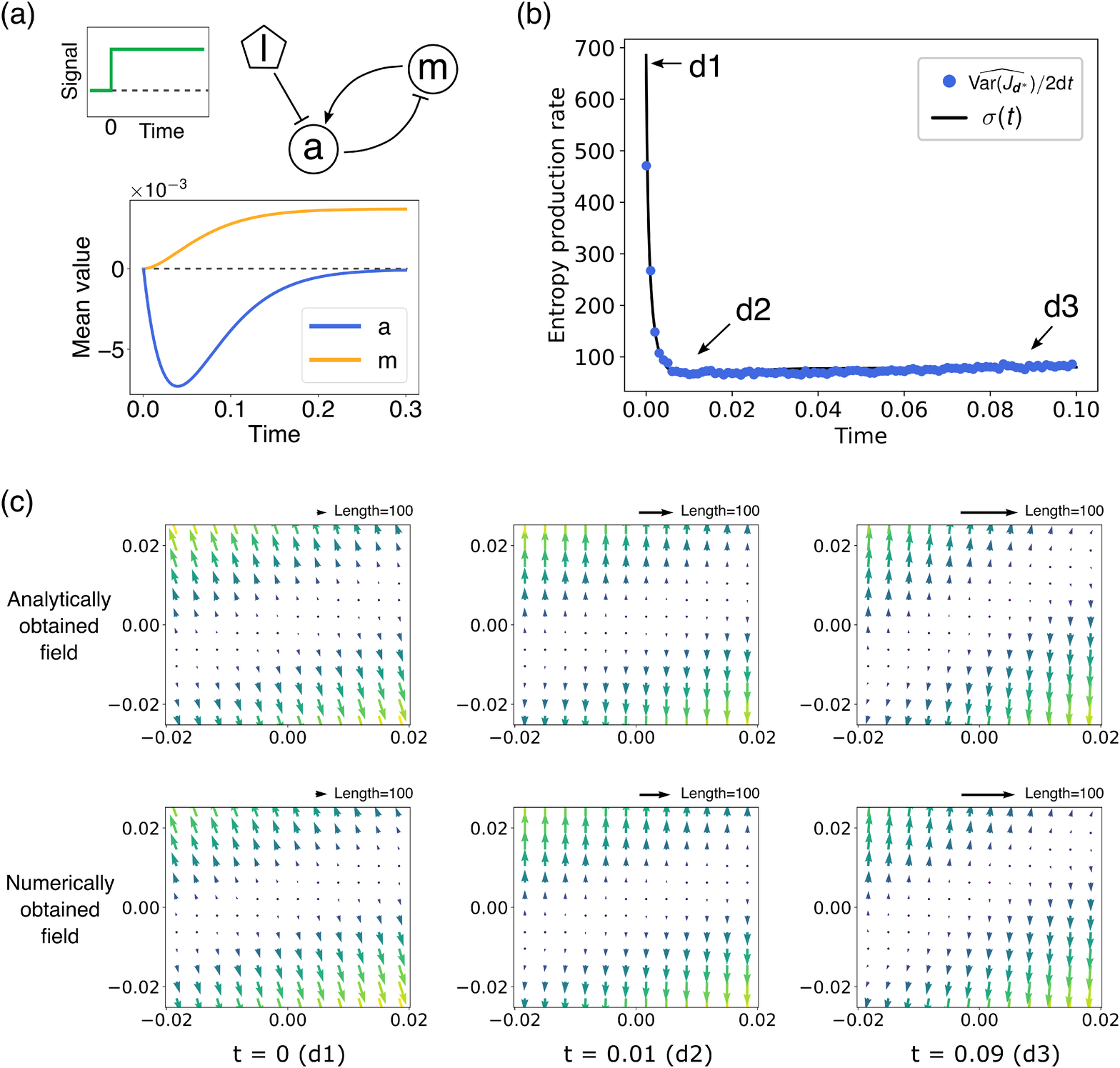}
\caption{Estimation in the adaptation model: (a) Sketch of the model. The average dynamics of $a$ and $m$ after the switching of the inhibitory input $l$ are plotted.
(b) Estimated entropy production rate. The blue circles are the estimated values \textcolor{black}{using the variance-based estimator (Eq.~(\ref{eq: variance-based})) after training the model function using $10^4$ trajectories}, and the black line is the true entropy production rate. \textcolor{black}{The estimated values are the results of a single trial.}
(c) Estimated thermodynamic force. The lower three figures are the estimated fields at $t=0$, $t = 0.01$, and $t = 0.09$  using \textcolor{black}{$10^4$} trajectories, and the upper three figures are the corresponding analytical solutions. Here the horizontal axis is the direction of $a$, the vertical axis is that of $m$, and an arrow with a length 100 is shown at the top of each figure. Note that in this particular case, the thermodynamic force becomes weaker as time evolves, and hence the magnitude of the vectors reduce.  For (a)-(c), the system parameters are set as $\tau_a = 0.02, \tau_m = 0.2, \alpha = 2.7, \beta = 1, \Delta_a = 0.005~(t<0), 0.5~(t\geq 0), \Delta_m = 0.005, l(t) = 0~(t<0), 0.01~(t\geq0)$, which are taken from realistic parameters of {\it E. coli} chemotaxis \cite{Tostevin2009, Ito2015}, the trajectories are generated with setting $\Delta t = 10^{-4}$, $\tau_{\rm obs} = 0.1$, and \textcolor{black}{the simple dual representation (Eq.~(\ref{eq: simple})) is used for training the model function.}
}
\label{fig: main2}
\end{figure*}
We demonstrate the effectiveness of our inference scheme with the following two linear Langevin models: (i) a one-dimensional breathing parabola model, and (ii) a two-dimensional adaptation model. In both models, non-stationary dynamics are repeatedly simulated with the same protocol, and a number of trajectories are sampled.
We estimate the entropy production rate solely on the basis of the trajectories, and compare the results with the analytical solutions (see the Supplementary Information for the analytical calculations).
Here, these linear models are adopted only to facilitate comparison with analytical solutions, and there is no hindrance to applying our method to nonlinear systems as well \cite{Shun:eem}. \\ \indent
We first consider the breathing parabola model that describes a one-dimensional colloidal system in a harmonic-trap $V(x, t) = \frac{\kappa(t)}{2}x^2$, \textcolor{black}{where $\kappa(t)$ is the time-dependent stiffness of the trap. This is a well-studied model in stochastic thermodynamics  \cite{bpexpt, opt,chvosta2020statistics} and has been used to experimentally realize microscopic heat engines consisting of a single colloidal particle as the working substance \cite{Blickle:2012rms,Martinez:2016bce} }. 
The dynamics can be accurately described by the following overdamped Langevin equation:
\begin{eqnarray}
\dot{x}(t) = -\frac{\kappa(t)}{\gamma}x(t) + \sqrt{2 D}\;\eta(t).\label{eq:BreathingP}
\end{eqnarray}
Here $\gamma$ is the viscous drag, and $\eta$ is a Gaussian white noise.
We consider the case that the system is initially in equilibrium and driven out of equilibrium as the potential changes with time. \textcolor{black}{Explicitly, we consider a protocol, $\kappa(t) = \gamma\alpha/(1+\alpha t)$, where the parameters $\alpha, \; \gamma$ as well as the diffusion constant $D$ are chosen such that they correspond to the experimental parameter set used in \cite{Blickle:2012rms} (see the Supplementary Information). 
}\\ \indent
In Fig. \ref{fig: main1}, we illustrate the central results of this paper for the breathing parabola model. We consider multiple realizations of the process of time duration $\tau_{\rm obs}$ as time series data (Fig.~\ref{fig: main1}(a)). The inference takes this as input and produces as output the entropy production at the level of an individual trajectory $\widehat{S}(t)$ for any single choice of realization (Fig.~\ref{fig: main1}(b)), as well as the average entropy production rate $\widehat{\sigma}(t)$ (Fig.~\ref{fig: main1}(c)). Here, the entropy production along a single trajectory $\widehat{S}(t)$ is estimated by summing up the estimated single-step entropy production:
\begin{eqnarray}
\Delta\widehat{S}(t) := {\bm d}\left(\frac{{\bm x}_t + {\bm x}_{t+\Delta t}}{2}, t + \frac{\Delta t}{2}~\bigg|{\bm \theta}^*\right)\; ({\bm x}_{t+\Delta t} - {\bm x}_t),~~ \label{eq: dS}
\end{eqnarray}
while the true entropy production $S(t)$ is calculated by summing up the true single-step entropy production:
\begin{eqnarray}
\Delta S(t) := {\bm F}\left(\frac{{\bm x}_t + {\bm x}_{t+\Delta t}}{2}, t + \frac{\Delta t}{2}\right)\; ({\bm x}_{t+\Delta t} - {\bm x}_t).\label{eq: dS_true}
\end{eqnarray}
Note that their dependence on the realization ${\bm x}(\cdot)$ is omitted in this notation for simplicity.\\ \indent
Specifically, we model the coefficient field ${\bm d}({\bm x}, t|{\bm\theta})$ by a feedforward neural network, and conduct the stochastic gradient ascent using an ensemble of single transitions extracted from $10^4$ or $10^5$ trajectories (see the Methods section for the details of the implementation) with $\Delta t = 10^{-2}~{\rm s}$ and $\tau_{\rm obs}=1~{\rm s}$. \textcolor{black}{ We note that, in recent experiments with colloidal systems, a few thousands of realizations of the trajectories have been realized with sampling intervals as small as $\Delta t = 10^{-6}~{\rm s}$ \cite{kumar2020exponentially}, and trajectory lengths as long as many tens of seconds \cite{Martinez:2016bce,Blickle:2012rms}.}\\ \indent
A feedforward neural network is adopted because it is suitable for expressing the non-trivial functional form of the thermodynamic force ${\bm F}({\bm x}, t)$ \cite{Hornik1989, kim2020learning}, and for continuous interpolation of discrete transition data \cite{Moon2019}. 
\textcolor{black}{In Fig.~\ref{fig: main1}(b), the entropy production is estimated along a single trajectory. We can confirm the good agreement with the analytical value. In Fig.~\ref{fig: main1}(c), the entropy production rate is estimated using $10^4$ and $10^5$ trajectories. In both the cases, the simple dual representation is used to train the model function on half the number of trajectories. On the other half, we use both the simple dual representation as well as the variance-based estimator in Eq.\ \eqref{eq: variance-based} for the estimation, in order to compare their relative merits. We see, quite surprisingly, that the variance-based estimator performs better than the simple dual representation and has much less statistical error. Since the simple dual representation is essentially just a weighted sum of the mean and variance, this implies that the error in it is due to the noise in the mean, as also explained above (and in the Supplementary Information).}\\ \indent
Another advantage of our method is that it also spatially resolves the thermodynamic force ${\bm F}({\bm x}, t)$, which would be hard to compute otherwise.
To demonstrate this point, we further analyze a two-dimensional model that has been used to study the adaptive behavior of living systems \cite{Tostevin2009, Lan2012, Ito2015,matsumoto2018role}. The model consists of the output activity $a$, the feedback controller $m$, and the input signal $l$, which we treat as a deterministic protocol. The dynamics of $a$ and $m$ are described by the following coupled Langevin equations:
\begin{subequations}
\label{eq:bioeq}
\begin{eqnarray}
\dot{a}(t) &=&-\frac{1}{\tau_a}\left[a(t) - \bar{a}(m(t), l(t))\right] + \sqrt{2\Delta_a}\;\eta_a(t),~~~~\\
\dot{m}(t) &=&-\frac{1}{\tau_m}a(t) + \sqrt{2\Delta_m}\;\eta_m(t),
\end{eqnarray}
\end{subequations}
where $\eta_a$ and $\eta_m$ are  independent Gaussian white noises, $\bar{a}(m(t), l(t))$ is the stationary value of $a$ given the instantaneous value of $m$ and $l$, and a linear function $\bar{a}(m(t), l(t)) = \alpha m(t) - \beta l(t)$ is adopted in this study.\\ \indent
We consider dynamics after the switching of the input as described in Fig.~\ref{fig: main2}(a). For a separation of time scales $\tau_m\gg\tau_a$,  the activity responds to the signal for a while before relaxing to a signal-independent value, which is called adaptation \cite{Lan2012}. Adaptation plays an important role in living systems for maintaining their sensitivity and fitness in time-varying environments. Specifically, this model studies {\it E. coli} chemotaxis \cite{Tostevin2009, Lan2012, Ito2015,matsumoto2018role} as an example. In this case, the activity regulates the motion of {\it E. coli} to move in the direction of higher concentration of input molecules by sensing the change in the concentration as described in Fig.~\ref{fig: main2}(a).\\ \indent
In this setup, the system is initially in a non-equilibrium stationary state (for $t<0$), and the signal change at $t = 0$ drives the system to a different non-equilibrium stationary state.
We show the results of the estimation of the entropy production rate and the thermodynamic force in Fig.~\ref{fig: main2}(b) and (c), respectively.
Because of the perturbation at $t = 0$, the non-equilibrium properties change sharply at the beginning.
Nonetheless, the model function ${\bm d}({\bm x}, t|{\bm \theta}^*)$ estimates the thermodynamic force well for the whole time interval (Fig.~\ref{fig: main2}(c)), and thus the entropy production rate as well (Fig.~\ref{fig: main2}(b)).
\textcolor{black}{In particular, we plot the result of a single trial in Fig.~\ref{fig: main2}(b), which means that the statistical error is negligible with only $10^4$ trajectories.} 
We note that the entropy production rate is orders of magnitude higher than that of the breathing parabola model. The results of Figs.~\ref{fig: main1} and \ref{fig: main2} demonstrate the effectiveness of our method in estimating a wide range of entropy production values accurately. \textcolor{black}{In the numerical experiments, we have used $\Delta t = 10^{-4}~{\rm s}$. We note that sampling resolutions in the range $\Delta t =10^{-6}~{\rm s}$ to $10^{-3}~{\rm s}$ have been shown to be feasible in realistic biological experiments \cite{monzel2016measuring}. We also note that an order of $10^3$ realizations are typical in DNA pulling experiments \cite{camunas2017experimental}}. \\ \indent
The thermodynamic force in Fig.~\ref{fig: main2}(c) has information about the spatial trend of the dynamics as well as the associated dissipation, since it is proportional to the mean local velocity ${\bm F}({\bm x}, t)\propto {\bm j}({\bm x}, t)/p({\bm x}, t)$ when the diffusion constant is homogeneous in space. At the beginning of the dynamics ($t = 0$), the state of the system tends to expand outside, reflecting the sudden increase of the noise intensity $\Delta_a$. Then, the stationary current around the distribution gradually emerges as the system relaxes to the new stationary state. Interestingly, the thermodynamic force aligns along the $m$-axis at $t = 0.01$, and thus the dynamics of $a$ becomes dissipationless.
The dissipation associated with the jumps of $a$ tends to be small for the whole time interval, which might have some biological implication as discussed in Refs.~\cite{Ito2015,matsumoto2018role}.\\ \indent
\begin{figure*}[t]
\includegraphics[width = 0.98\linewidth]{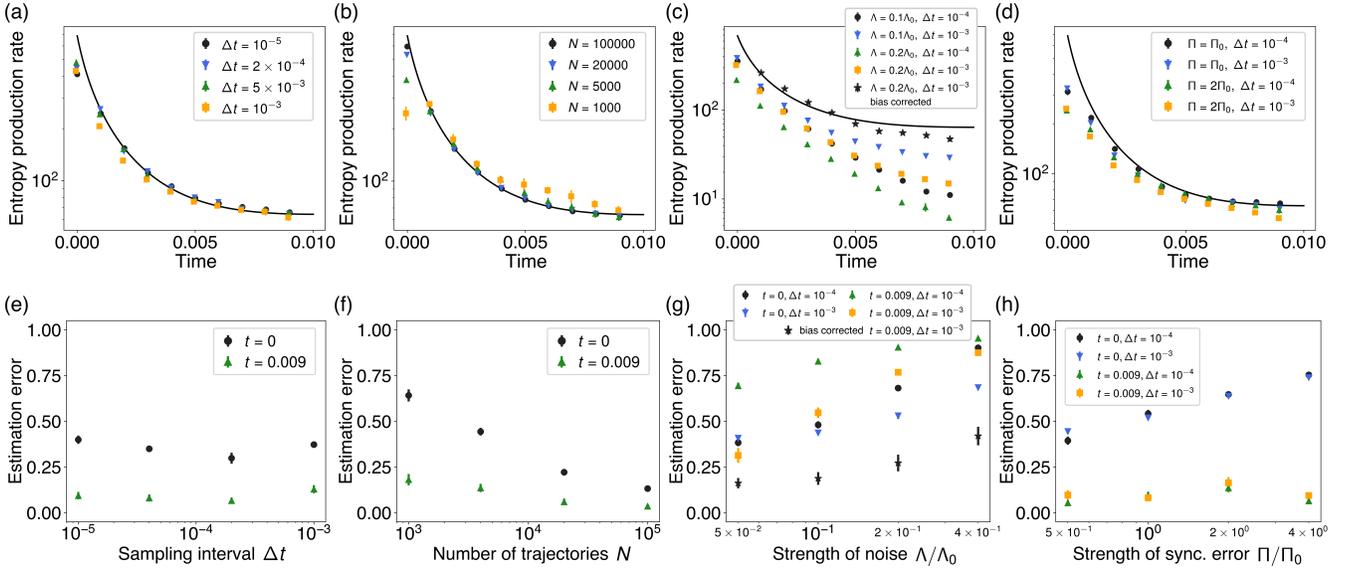}
\caption{
\textcolor{black}{
Estimation with imperfect data in the adaptation model:
(a) with different sampling intervals, 
(b) with different number of trajectories,
(c) with the measurement noise (Eq.~(\ref{eq: noise})), 
and (d) with the synchronization error (Eq.~(\ref{eq: sync})).
(e)-(h) is the estimation error ($:=|\widehat{\sigma}(t)-\sigma(t)|/\sigma(t)$) analysis of (a)-(d).
(a)(e) The estimation is robust against the choice of the sampling interval. (b)(f) The number of trajectories required for the convergence is large near the initial time when the dynamics is highly non-stationary. (c)(g) As the strength of the measurement noise $\Lambda$ increases, the estimate becomes lower than the true value because the measurement noise effectively increases the diffusion matrix. A larger time interval for the generalized current can mitigate this effect, and the bias-corrected estimation (Eq.~(\ref{eq: debias})) substantially reduces the bias. Note that the estimate at $t=0$ is not debiased, since the bias-corrected estimator of the variance requires single transition data before the target time (see Eq.~(\ref{eq: debias})).
(d)(h) The estimate becomes an averaged value in the time direction. In contrast to (c)(g), the time interval dependence is small.
For (a)-(h), the simple dual representation (Eq.~(\ref{eq: simple})) is used for training the model function, and the variance-based estimator (Eq.~(\ref{eq: variance-based})) is used for the estimation. The mean and its standard deviation of ten independent trials are plotted.
The system parameters are the same as those in Fig.~\ref{fig: main2} except $\tau_{\rm obs} = 0.01$.
}}
\label{fig: main3}
\end{figure*}
\textcolor{black}{So far, we have shown that our inference scheme estimates the entropy production very well in  ideal data sets. Next, we demonstrate the practical effectiveness of our algorithm by considering the dependence of the inference scheme on (i) the sampling interval, (ii) the number of trajectories, (iii) measurement noise, and (iv) time-synchronization error. The analysis is carried out in the adaptation model, for times $t = 0$ and $t=0.009$, at which the degrees of non-stationarity are different. The results are summarized in Fig.\ \ref{fig: main3}. In most of the cases, we find that the estimation error defined by $\left|\widehat{\sigma}(t) - \sigma(t)\right|/\sigma(t)$ is higher at $t=0$ when the system is highly non-stationary.\\ \indent
In Fig.\ \ref{fig: main3}(a) and (e), we demonstrate the effect of the sampling interval $\Delta t$ on the estimation.  For both the $t$ values, we find that the estimation error does not significantly depend on the sampling interval $\Delta t$ in the range $10^{-5}$ to $10^{-3}$, which demonstrates the robustness of our method against $\Delta t$.\\ \indent  
In Fig.~\ref{fig: main3}(b) and (f), we consider the dependence of the estimated entropy production rate on $N$ - the number of trajectories used for the estimation. We find that roughly $10^3$ trajectories are required to get an estimate that is within $0.25$ error of the true value for $t=0.009$. On the other hand, we need at least $10^4$ trajectories at $t=0$ to get an estimate within the same accuracy. 
This is because the system is highly non-stationary at $t=0$ and thus the benefit of the continuous-time inference decreases.\\ \indent
In Fig.~\ref{fig: main3}(c) and (g), the effect of measurement noise is studied. Here, the measurement noise is added to trajectory data as follows:
\begin{eqnarray}
{\bm y}_{j\Delta t} = {\bm x}_{j\Delta t} + \sqrt{\Lambda}{\bm \eta^j},\label{eq: noise}
\end{eqnarray}
where $\Lambda$ is the strength of the noise, and $\eta$ is a Gaussian white noise satisfying $\left<\eta^i_a\eta^j_b\right> = \delta_{a, b}\delta_{i, j}$.
The strength $\Lambda$ is compared to $\Lambda_0 = 0.03$ which is around the standard deviation of the variable $m$ in the stationary state at $t > 0$.
We find that the estimate becomes lower in value as the strength $\Lambda$ increases, while a larger time interval for the generalized current can mitigate this effect.
This result can be explained by the fact that the measurement noise effectively increases the diffusion matrix, and its effect becomes small as $\Delta t$ increases since the Langevin noise scales as $\propto \sqrt{{\Delta}t}$ while the contribution from the measurement noise is independent of $\Delta t$. Since the bias in $\widehat{{\rm Var}(J_{\bm d})}$ is the major source of the estimation error, we expect that the use of a bias-corrected estimator \cite{Vestergaard2014, Frishman2020} will reduce this error. Indeed, we do find that the bias-corrected estimator, star symbols in Fig.~\ref{fig: main3}(c) and (g), significantly reduces the estimation error (see the Methods section for the details). \\ \indent
Finally in Fig.~\ref{fig: main3}(d) and (h), the effect of synchronization error is studied.
We introduce the synchronization error by starting the sampling of each trajectory at $\tilde{t}$ and regarding the sampled trajectories as the states at $t = 0, \Delta t, 2\Delta t, ...$ (actual time series is $t = \tilde{t}, \tilde{t}+\Delta t, ...$). Here, $\tilde{t}$ is a stochastic variable defined by
\begin{eqnarray}
\tilde{t} = \left\lfloor\frac{{\rm uni}(0, \Pi)}{\Delta t'}\right\rfloor\Delta t',\label{eq: sync}
\end{eqnarray}
where ${\rm uni}(0, \Pi)$ returns the value $x$ uniformly randomly from $0<x<\Pi$, the brackets are the floor function, and $\Delta t' = 10^{-4}$ is used independent of $\Delta t$. 
The strength $\Pi$ is compared to $\Pi_0$ which approximately satisfies $\sigma(\Pi_0)\approx \sigma(0)/2$.
We find that the estimate becomes an averaged value in the time direction, and the time interval dependence is small in this case.\\ \indent
In conclusion we find that our inference scheme is robust to deviations from an ideal dataset for experimentally feasible parameter values and even steep rates of change of the entropy production over short time intervals. 
}\\ \indent

\section*{Discussion}
The main contribution of this work is the insight that variational schemes can be used to estimate the exact entropy production rate of a non-stationary system under arbitrary conditions, given the constraints of Markovianity. The different variational representations of the entropy production rate: $\sigma_{\rm NEEP}$, $\sigma_{\rm Simple}$ and $\sigma_{\rm TUR}$, as well as their close relation to each other, are clarified in terms of the range of applicability, the optimal coefficient field and the tightness of the bound in each case, as summarized in Table~\ref{table: summary}.  \\ \indent
Our second main contribution is the algorithm we develop to implement the variational schemes, by means of continuous-time inference, namely
using the constraint that $d^*$ has to be continuous in time, to infer it in one shot for the full time range of interest. 
\textcolor{black}{In addition, we find that the variance-based estimator of the entropy production rate, performs significantly better than other estimators, in the case when our algorithm is optimised to take full advantage of the continuous-time inference. We expect that this  property will be of great practical use in estimating entropy production for non-stationary systems.}
The continuous-time inference is enabled by the high representation ability of the neural network, and can be implemented without any prior assumptions on the functional form of the thermodynamic force ${\bm F}({\bm x}, t)$. 
Our work shows that the neural network can effectively learn the field even if it is time-dependent, thus opening up a wide range of possibilities for future applications.\\ \indent
Our studies regarding the practical effectiveness of our scheme when applied to data that might conceivably contain one of several sources of noise, indicate that these tools could also be applied to the study of biological \cite{battle2016broken} or active matter systems \cite{Ramaswamy:MSA}. \textcolor{black}{It will be very interesting to see whether these results can be used to infer new information from the empirical data we have from molecular motors such as kinesin \cite{schnitzer1997kinesin} or ${\rm F}_1$-ATPase \cite{duncan1995rotation,toyabe2013experimental}}. The thermodynamics of cooling or warming up in classical systems \cite{Alessio} or the study of quantum systems being monitored by a sequence of measurements \cite{Hekking2013, Horowitz2013, dressel2017arrow,manikandan2019fluctuation} are other promising areas to which these results can be applied.

\section*{Methods}
\begin{algorithm}[t]
\caption{Train the model function ${\bm d}({\bm x}, t|{\bm\theta})$}
\label{algorithm}
\begin{algorithmic}
\REQUIRE $N$ trajectories with length $M\Delta t$  $$\Gamma^{(i)} = \left\{{\bm x}^{(i)}_{0}, {\bm x}^{(i)}_{\Delta t}, ..., {\bm x}^{(i)}_{M\Delta t} \right\}~(i = 1, ..., N)$$
\STATE Normalize the data as $$x^{(i)}_{j\Delta t}\leftarrow \left(x^{(i)}_{j\Delta t}-{\rm mean}\left[x\right]\right)/{\rm std\left[x\right]}$$
\LOOP
\STATE Choose a hyperparameter set $\{n_{\rm layer}, n_{\rm hidden}, n_{\rm output}\}$
\STATE ${\bm \theta}\leftarrow$ initialize network parameters
\LOOP
\FOR{$j = 0, 1, ..., M-1$}
\STATE
Compute the current using $\Gamma^{(i)} ~(i=1,..., N/2)$
\scriptsize
\begin{eqnarray*}
J_{\bm d}^{(i)} = {\bm d}\left(\frac{{\bm x}^{(i)}_{j\Delta t} + {\bm x}^{(i)}_{(j+1)\Delta t}}{2}, \left[j +\frac{1}{2}\right]\Delta t ~\Bigg| {\bm \theta}\right) ({\bm x}^{(i)}_{(j+1)\Delta t} - {\bm x}^{(i)}_{j\Delta t})
\end{eqnarray*}
\normalsize
\STATE
Compute $\widehat{\sigma}(j\Delta t)$ using $\{J_{\bm d}^{(i)}\}_{i=1, ..., N/2}$
\ENDFOR
\STATE
Compute the objective function
$$\widehat{f}({\bm \theta})|_{\rm train} = \frac{1}{M}\sum_{j=0}^{M-1} \widehat{\sigma}(j\Delta t)$$
Update the parameters
$${\bm \theta}\leftarrow {\bm\theta} + \alpha\partial_{\bm \theta}\widehat{f}|_{\rm train}$$
Compute the test value $\widehat{f}({\bm \theta})|_{\rm test}$
using $\Gamma^{(i)}~(i=N/2+1,..., N)$ in the same manner
\STATE
Record $\widehat{f}({\bm \theta})|_{\rm test}$, ${\bm\theta}$, and the hyperparameter set
\ENDLOOP
\ENDLOOP
\STATE
${\bm\theta}^*\leftarrow$ parameters that maximize $\widehat{f}({\bm \theta})|_{\rm test}$ in the record
\end{algorithmic}
\end{algorithm}

\begin{figure}[t]
\includegraphics[width = \linewidth]{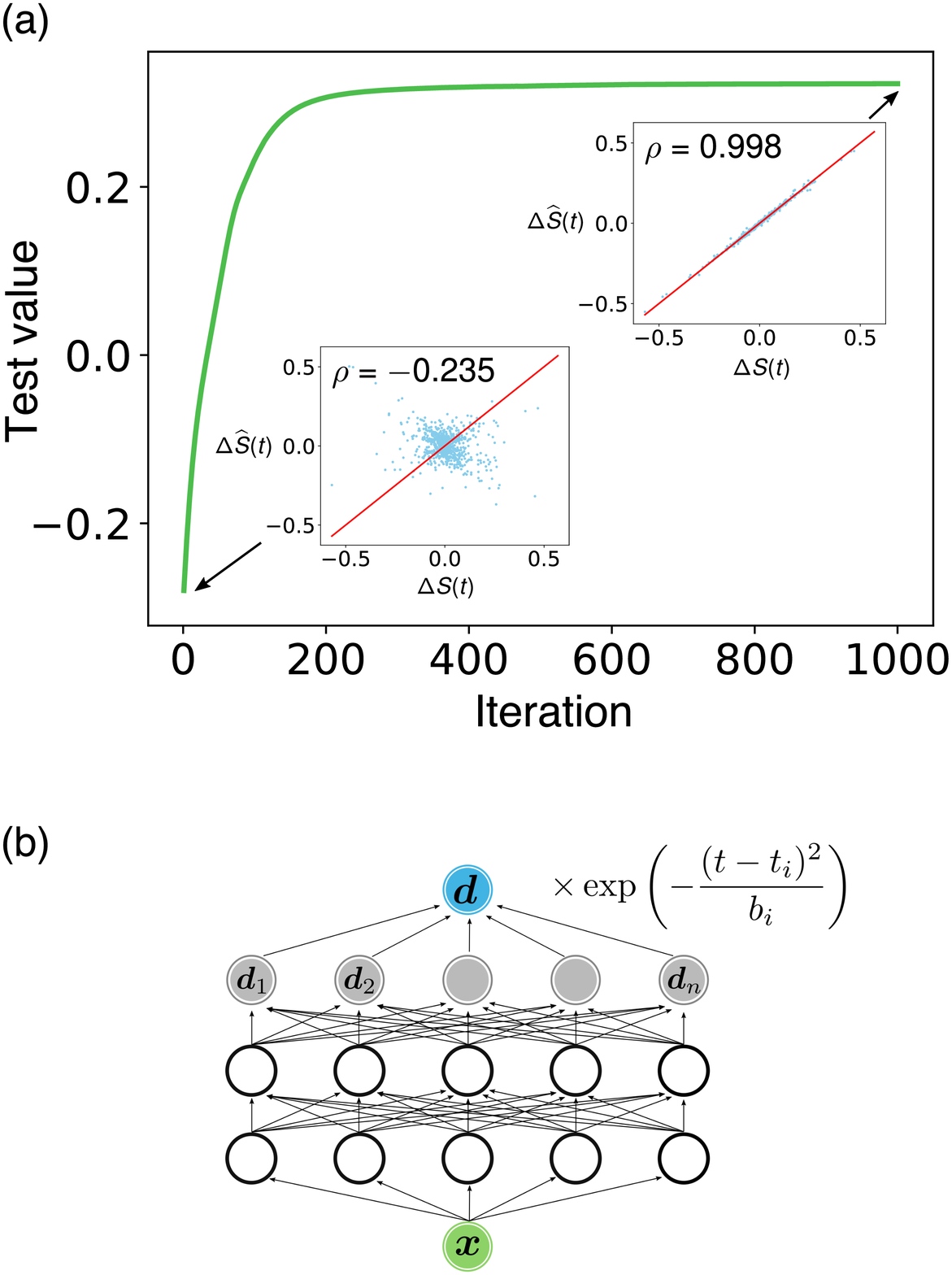}
\caption{Training of the neural network: (a) Example of the learning curve for the breathing parabola model with the same system parameters as in Fig.~\ref{fig: main1}. We show scatter plots between the estimated (Eq.~(\ref{eq: dS})) and the true (Eq.~(\ref{eq: dS_true})) single-step entropy production along a single trajectory as inset figures, and $\rho$ is the correlation between them evaluated using 10 trajectories. As the gradient ascent updates the parameters of the neural network, the estimates of the entropy production become accurate.
(b) Sketch of the neural network. \textcolor{black}{We depict a four-layer network with two hidden layers as an example.} The output layer consists of the gray and blue units, which give the output value as Eq.~(\ref{eq: modified}).
}
\label{fig: supply1}
\end{figure}

\noindent
\textbf{Model function and gradient ascent.}
In Algorithm~\ref{algorithm}, we present the overall algorithm of our estimation method.
In the following, we explain its implementation in more detail.\\ \indent
\textcolor{black}{
The algorithm takes $N$ trajectories with length $M\Delta t$ as input, and outputs the optimal parameters of the model function. First of all, we normalize the trajectory data as
\begin{eqnarray}
x^{(i)}_{j\Delta t}\leftarrow \left(x^{(i)}_{j\Delta t}-{\rm mean}\left[x\right]\right)/{\rm std\left[x\right]},\label{eq: normalization}
\end{eqnarray}
where ${\rm mean}\left[x\right]$ and ${\rm std\left[x\right]}$ are the mean and the standard deviation of the position ${\bm x}^{(i)}_{j\Delta t}$ for all $i$ and $j$. We find that the normalization improves the convergence of the gradient ascent (see the Supplementary Information for the numerical results).\\ \indent
}
Next, we explain the main part of the algorithm between the inner loop of Algorithm~\ref{algorithm}.
\textcolor{black}{In this study, we adopt a multi-layer feedforward network depicted in Fig.~\ref{fig: supply1}(b) to model the time-dependent coefficient field ${\bm d}({\bm x}, t|{\bm\theta})$.
The output of each layer is fully-connected to the next layer, and the rectified linear function (ReLU) is adopted as the activation function except at the output layer.
The output is a linear combination of the output layer using Gaussian functions:
\begin{eqnarray}
{\bm d} = \sum_{i=1}^{N_{\rm output}}{\bm d}_i\exp\left[-\left(\frac{t-t_i}{b_i}\right)^2\right],\label{eq: modified}
\end{eqnarray}
where $N_{\rm output}$ is the number of units in the output layer, and ${\bm d}_i$ is the vector value of the unit, which depends on ${\bm x}$.}
The centers and widths of the Gaussian functions $t_i$ and $b_i$ are parameters to optimize, which are initialized by $t_i = (i-1)\tau_{\rm obs}/(N_{\rm output}-1)$ and $b_i = \tau_{\rm obs}/(N_{\rm output} -1)$.
The idea of this network is that the unit ${\bm d}_i$ learns the thermodynamic force around time $t_i$.
\textcolor{black}{Note that the number of hidden layers $N_{\rm layer}$, the number of units per hidden layer $N_{\rm hidden}$, and $N_{\rm output}$ are the hyperparameters which should be determined before the gradient ascent. The process of determining hyperparameters is explained in the next section.}\\ \indent
We conduct the gradient ascent with respect to the parameters ${\bm \theta}$ of the neural network using the objective function (\ref{eq: objective_func}).
The ensemble of single transitions $\left\{{\bm x}^{(i)}_{j\Delta t}, {\bm x}^{(i)}_{(j+1)\Delta t}\right\}$ is used to calculate $\widehat{\sigma}_{\rm Simple}(j\Delta t) = \frac{1}{\Delta t}\left[2\widehat{\left<J_{\bm d}\right>}-\widehat{\frac{{\rm Var}(J_{\bm d})}{2}}\right]$ by regarding each ${\bm d}\left(\frac{{\bm x}^{(i)}_{j\Delta t} + {\bm x}^{(i)}_{(j+1)\Delta t}}{2}, \left(j+\frac{1}{2}\right)\Delta t ~\big| {\bm \theta}\right) \left({\bm x}^{(i)}_{(j+1)\Delta t} - {\bm x}^{(i)}_{j\Delta t}\right)$ as a realization of the generalized current $J_{\bm d}$ at time $j\Delta t$. Here, the simple dual representation is used as an example for the explanation. Then, the basic update rule of the gradient ascent is as follows: 
\begin{eqnarray}
{\bm\theta}\leftarrow {\bm\theta} + \alpha\partial_{\bm \theta}\widehat{f},
\end{eqnarray}
where $\alpha$ is the step size, and $\widehat{f}$ is the estimated objective function defined in Eq.~(\ref{eq: objective_func}). Since the parameters are updated towards the direction in which the objective function increases the most, ${\bm d}({\bm x}, t|{\bm \theta})$ gets close to the thermodynamic force, and the estimates of the entropy production become accurate as shown in Fig.~\ref{fig: supply1}(a). Specifically, we implement an algorithm called Adam \cite{Kingma2014} for the gradient ascent to improve the convergence.
The hyperparameters of Adam are set to the values suggested in \cite{Kingma2014}, for example $\alpha=10^{-3}$.\\ \indent
\begin{figure*}
\includegraphics[width = \linewidth]{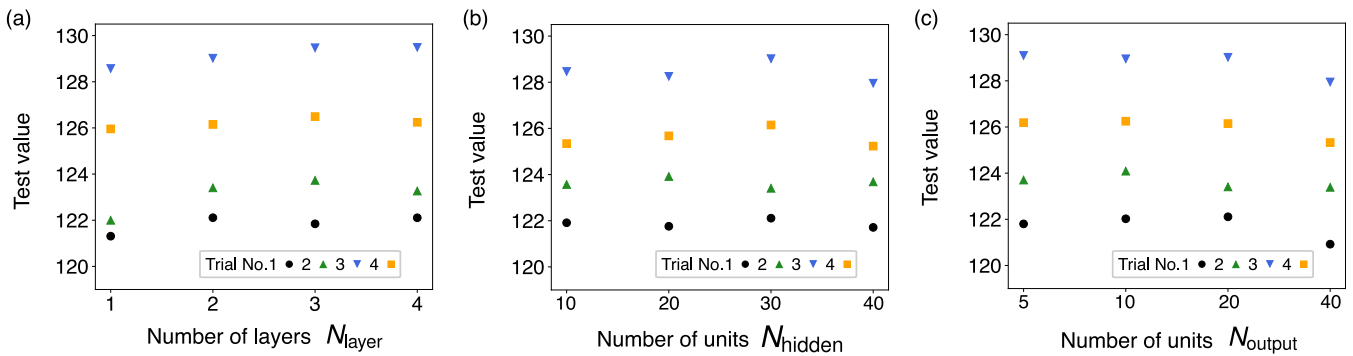}
\caption{Hyperparameter tuning of the modified network in the adaptation model with (a) $N_{\rm hidden}=30$, $N_{\rm output}=20$, (b) $N_{\rm layer}=2$, $N_{\rm output}=20$, and (c) $N_{\rm layer}=2$, $N_{\rm hidden}=30$. The test values of four independent trials are plotted with different markers respectively. In practice, the results of a single marker can be obtained. For example, we can judge that underfitting occurs at $N_{\rm layer} = 1$, \textcolor{black}{since the test value is small compared to the other cases.} The system parameters of the adaptation model are the same as those in Fig.~\ref{fig: main3}, and $10^4$ trajectories are used for each trial.}
\label{fig: hyper}
\end{figure*}
\begin{table*}
\begin{center}
\begin{tabular}{c | c c c c c c} \hline
Model &	$N_{\rm traj}$ & $N_{\rm layer}$	&$N_{\rm hidden}$	&$N_{\rm output}$	&$N_{\rm step}$	& $N_{\rm param}$ \\ \hline
Breathing parabola model  & $10^4$  &  3    & 10    & 5      & 1000		& 305	\\
                          & $10^5$  &  3    & 10    & 10     & 1000	    & 370   \\ \hline
Adaptation model          & $10^3-10^5$  & 2    & 30   & 20	 & 5000	    & 2300  \\ \hline
\end{tabular}
\caption{Settings of the neural network.
\textcolor{black}{$N_{\rm traj}$ is the number of trajectories used for training the neural network.}
$N_{\rm layer}, N_{\rm hidden}$ and $N_{\rm output}$ are the hyperparameters that are determined by hand.
$N_{\rm step}$ is the iteration number of the gradient ascent. $N_{\rm param}$ is the total number of parameters to optimize by the gradient ascent. \textcolor{black}{We use the same hyperparameter set for the adaptation model independent of $N_{\rm traj}$, since the variation of $N_{\rm traj}$ is only studied in Fig.~\ref{fig: main3}(b) and (f) and the purpose of these figures is to show the effect of smaller ensemble sizes on the estimation.}
}
\label{table: hyper2}
\end{center}
\end{table*}
\textcolor{black}{
Finally, we remark on the bias-corrected estimator of the variance in the presence of  measurement noise.
Here, we use ${\bm y}^{(i)}_{j\Delta t}$ for the observed position, which is the sum of the true position ${\bm x}^{(i)}_{j\Delta t}$ and the measurement noise as defined in Eq.~(\ref{eq: noise}). Then, the standard estimator of the variance using the generalized current $J_{\bm d}(j\Delta t)|_{\bm y} = {\bm d}\left(\frac{{\bm y}^{(i)}_{j\Delta t} + {\bm y}^{(i)}_{(j+1)\Delta t}}{2}, \left(j+\frac{1}{2}\right)\Delta t ~\big| {\bm \theta}\right) \left({\bm y}^{(i)}_{(j+1)\Delta t} - {\bm y}^{(i)}_{j\Delta t}\right)$ is shown to be severely biased in the sense that the term with $\Lambda$ becomes the leading order term in $\Delta t$:
\begin{eqnarray}
&&\widehat{{\rm Var}(J_{\bm d})} =
\left<\left(J_{\bm d}(j\Delta t)|_{\bm y} - \left<J_{\bm d}(j\Delta t)|_{\bm y}\right>\right)^2\right>\nonumber\\
&&~~=\left<\left(J_{\bm d}(j\Delta t)|_{\bm x} - \left<J_{\bm d}(j\Delta t)|_{\bm x}\right>\right)^2\right> + 2\Lambda\left<{\bm d}^2\right> + O(\Delta t)\nonumber\\
&&~~=2\Lambda\left<{\bm d}^2\right> + O(\Delta t),
\end{eqnarray}
where $\left<\cdots\right>$ denotes the average over the index $i$, and $|_{\bm x}$ or $|_{\bm y}$ is added to clarify which variable is used to calculate the current. Fortunately, this bias can be corrected by modifying the expression in Refs.~\cite{Vestergaard2014, Frishman2020} as follows:
\begin{eqnarray}
\widehat{{\rm Var}(J_{\bm d})} &=& \frac{1}{2}\left<\left\{J_{\bm d}(j\Delta t) + J_{\bm d}((j-1)\Delta t)\right\}^2\right>\nonumber\\
&&+\left<J_{\bm d}(j\Delta t)J_{\bm d}((j-1)\Delta t)\right>
,\label{eq: debias}
\end{eqnarray}
which gives an estimate of the variance at time $j\Delta t$, and cancels the $\Lambda\left<{\bm d}^2\right>$ term between the first and the second terms. This bias-corrected estimator is adopted in Fig.~\ref{fig: main3}(c) and (g), and shown to be very effective against the measurement noise.
We note that such a bias does not appear in the mean current $\widehat{\left<J_{\bm d}\right>}$ \cite{Frishman2020}.\\
}

\noindent
\textbf{Data splitting scheme.} 
As is the case for many machine learning problems, we should be careful about the problem of underfitting or overfitting.
To avoid such problems, we use the data splitting scheme (or cross validation) \cite{Shun:eem, kim2020learning}. Concretely, we use half of the ensemble of trajectories as the training data, and the other half as the test data.
The model function is trained by using only the training data, and the progress of learning is evaluated by using the test data.
When the ensemble size is small, there appears a maximum in the middle of the learning curve of the test value, which is a sign that, after this, the model function is overfitted to the training data.
Thus, we adopt the parameters at the peak, ${\bm\theta}^*$, for the estimation \cite{Amari1997}. \\ \indent
The data splitting scheme is also useful to determine the hyperparameters of the neural network.
The neural network that maximizes \textcolor{black}{the test value at ${\bm\theta}^*$ (we omit ``at ${\bm\theta}^*$" hereafter for simplicity)}, which is the mean of the output estimates \textcolor{black}{$\sum_{i}\widehat{\sigma}(i\Delta t)/M$}, would be the best since the test value typically behaves as follows: (i) When the model complexity is too low, both the test and training values become much smaller than the true entropy production rate since the model cannot express the thermodynamic force well. (ii) As we increase the model complexity, both the test and training values increase, and at some point, the test value reaches its maximum, often being closest to the true value from below. (iii) When the model complexity is too high, the training value becomes much bigger than the true value, while the test value becomes much smaller, due to overfitting. Thus, we adopt the hyperparameters that realize the highest test value for the estimation, and this process corresponds to the outer loop in Algorithm \ref{algorithm}.
This strategy is effective in practice since we just need to check the peak of the learning curve while trying several hyperparameters, and do not need  to know the true value of the entropy production rate (a similar discussion can be found in Refs.~\cite{Shun:eem,kim2020learning}). Note that the same variational representation should be used for both the training and evaluation.\\ \indent
In Fig.~\ref{fig: hyper}, we show an example of the hyperparameter tuning for the modified network in the adaptation model.
In Fig.~\ref{fig: hyper}(a)-(c), the hyperparameter dependence of the test value is shown for four independent trials. From these plots, for example, we can judge that the network complexity is not enough with $N_{\rm layer}=1$, while the networks with $N_{\rm layer}$ bigger than one show similar performance in Fig.~\ref{fig: hyper}(a). 
\textcolor{black}{Similarly, we can judge that the network complexity is too high with $N_{\rm output}=40$ in Fig.~\ref{fig: hyper}(c).}
In Table~\ref{table: hyper2}, the hyperparameter values used in our numerical experiments are summarized.

\section*{Code Availability}
Computer codes implementing our algorithm and interactive demo programs are available online at https://github.com/tsuboshun/LearnEntropy.

\section*{Data Availability}
Trajectory data used for testing our method are available from the corresponding author upon  request.

\begin{acknowledgements}
T. S. is supported by JSPS KAKENHI Grant No. 16H02211 and 19H05796.
T. S. is also supported by Institute of AI and Beyond of the University of Tokyo.
\end{acknowledgements}

\newpage
\begin{widetext}
\vspace{10mm}
\section*{Supplementary Information}
We provide supplementary information for the analytical and numerical calculations presented in the main text.

\subsection*{Supplementary Note 1: Details of the variational representations}
Here we discuss details of the variational representations such as their derivation and comparisons.
\subsubsection*{Derivation of the NEEP representation and its use in Markov jump processes}
Here we derive $\sigma_{\rm NEEP}$ on the basis of a dual representation of the Kullback-Leibler (KL) divergence, and clarify its applicability to non-stationary dynamics as well as its relation to the dual representation. In this subsection, we mainly consider Markov jump processes, and also discuss how to adapt our algorithm to this case.\\ \indent
We consider discrete probability distributions defined on the state space $\Omega$.
The KL divergence between probability distributions $P$ and $Q$ is defined as
\begin{eqnarray}
D_{\rm KL}(P || Q) := \sum_{x\in\Omega} P(x)\ln\frac{P(x)}{Q(x)}.
\end{eqnarray}
The KL divergence admits the following variational representation \cite{Keziou2003, Nguyen2010, Belghazi2018}:
\begin{eqnarray}
D_{\rm KL}(P || Q) = \max_{h\in\mathcal{F}}~ \mathbb{E}_P[h+1] - \mathbb{E}_Q[e^{h}],\label{eq: dual_KL}
\end{eqnarray}
where $\mathcal{F}$ is a set of functions $h: \Omega\rightarrow \mathbb{R}$ such that the two expectations are finite, and the optimal function is given by $h^*(x) = \ln\frac{P(x)}{Q(x)}$.
This is derived using the Fenchel convex duality \cite{Rockafellar1970, Nguyen2010}, and we call it a dual representation of the KL divergence.\\ \indent
We use the KL divergence formula for the entropy production rate \cite{Seifert:2012stf} as
\begin{eqnarray}
\sigma(t) = D_{\rm KL}(p_t(x)r_t(x, x') || p_t(x')r_t(x', x)),\label{eq: sigma}
\end{eqnarray}
where $p_t(x)$ is the probability distribution of the state and $r_t(x, x')$ is the transition rate from $x$ to $x'$ at time $t$. Then, we apply Eq.~(\ref{eq: dual_KL}) to Eq.~(\ref{eq: sigma}) to get a dual representation of the entropy production rate:
\begin{eqnarray}
\sigma(t) = \frac{1}{{\rm d}t}\max_{h\in\mathcal{F}'} \left<h - e^{-h} + 1\right>,\label{eq: dual_ep}
\end{eqnarray}
where $\mathcal{F}'$ is a set of functions $h: \Omega\times \Omega\rightarrow \mathbb{R}$ such that $h(x', x) = -h(x, x')$ and the above expectation is finite and calculated as
\begin{eqnarray}
\left<f(x, x') \right> :={\rm d}t\sum_{x, x'}p_t(x)r_t(x, x') f(x, x').
\end{eqnarray}
Here ${\rm d}t$ is added so that the expression becomes the same as in the main text in which the expectation is taken with respect to the joint probability distribution $p(x(t), x(t+{\rm d}t)) = p_t(x(t))r_t(x(t), x(t+{\rm d}t)){\rm d}t$.
The optimal function is given by $h^*(x, x') = \ln\frac{p_t(x)r_t(x, x')}{p_t(x')r_t(x', x)} $, which is the entropy production ${\rm d}S(x, x')$ associated with the jump. We note that Eq.~(\ref{eq: dual_ep}) holds for dynamics that satisfy ${\rm d}S(x', x) = -{\rm d}S(x, x')$ including Markov jump processes and overdamped Langevin dynamics. The dual representation (\ref{eq: dual_ep}) is equivalent to $\sigma_{\rm NEEP}$ defined in Eq.~(\ref{eq: NEEP}) of the main text if Langevin dynamics is considered.
Since nothing is assumed on $p_t(x)$ in Eq.~(\ref{eq: sigma}), $\sigma_{\rm NEEP}$ gives the exact entropy production rate even for non-stationary dynamics.\\ \indent
The derivation of Eq.~(\ref{eq: dual_ep}) is as follows:
\begin{subequations}
\begin{eqnarray}
\sigma &=& D_{\rm KL}(p_t(x)r_t(x, x') || p_t(x')r_t(x', x))\\
&=& \max_{h\in\mathcal{F}}\left[\sum_{x, x'} p_t(x)r_t(x, x')h(x, x') - \sum_{x, x'} p_t(x')r_t(x', x)e^{h(x, x')} + 1\right] \label{eq: dual_ep1}\\
&=& \max_{h\in\mathcal{F}'}\left[\sum_{x, x'} p_t(x)r_t(x, x')h(x, x') - \sum_{x, x'} p_t(x')r_t(x', x)e^{h(x, x')} + 1\right] \label{eq: dual_ep2}\\
&=& \max_{h\in\mathcal{F}'}\left[\sum_{x, x'} p_t(x)r_t(x, x')\left\{h(x, x') - e^{-h(x, x')} + 1\right\}\right]\\
&=& \frac{1}{{\rm d}t}\max_{h\in\mathcal{F}'}\left<h - e^{-h} + 1\right>,
\end{eqnarray}
\end{subequations}
where Eq.~(\ref{eq: dual_KL}) is used in Eq.~(\ref{eq: dual_ep1}), and a constraint $h(x', x) = -h(x, x')$ is newly added in Eq.~(\ref{eq: dual_ep2})
using the fact that the optimal function $h^*$ satisfies the constraint.\\ \indent
Next, we discuss the numerical estimation in Markov jump processes using $\sigma_{\rm NEEP}$.
As is the case in Langevin dynamics, we want to estimate the entropy production rate solely on the basis of an ensemble of trajectories sampled from a stochastic jump process with interval $\Delta t$:
\begin{eqnarray}
\Gamma_i = \left\{x_0, x_{\Delta t}, ..., x_{\tau_{\rm obs}} (= x_{M\Delta t})\right\}_i ~(i = 1, ..., N).
\end{eqnarray}
In general, it is necessary to reconstruct the underlying jump dynamics which occur between the sampling times, to obtain the exact estimate \cite{Shun:eem}, but here for simplicity, we consider the case that $\Delta t$ is sufficiently small so that transitions occur at most once between the sampling times.\\ \indent
The estimation procedure is almost the same as in Langevin dynamics.
We use the ensemble of single transitions $\{x_t, x_{t+\Delta t}\}$ to calculate $\widehat{\sigma}_{\rm NEEP}(t) = \frac{1}{\Delta t}\widehat{\left<h - e^{-h} + 1\right>}$ by regarding each $h\left(x_t, x_{t+\Delta t}, t+\frac{\Delta t}{2} | {\bm\theta}\right)\; (1-\delta_{x_t, x_{t+\Delta t}})$ as a realization of the generalized current. Here, $h(x, x', t| {\bm \theta})$ is a parametric model function that satisfies $h(x', x, t|{\bm \theta}) = -h(x, x', t|{\bm\theta})$, and we optimize the parameters ${\bm \theta}$ by the gradient ascent using the objective function defined in Eq.~(\ref{eq: objective_func}).\\ \indent
However, in contrast to the case of Langevin dynamics, a neural network may not be appropriate for the parametric model function because of the discreteness of the arguments $x$ and $x'$. This problem is addressed by transforming discrete states into continuous vectors with an embedding layer, as in Ref.~\cite{kim2020learning}.
Another way to define the model function would be to assign an independent parametric function $f_{x, x'}(t | {\bm \theta}_{x, x'})$ for each transition edge, and define the function $h$ as $h(x, x', t |{\bm \theta}) = \sum_{y, y'}f_{y, y'}(t | {\bm \theta}_{y, y'})\delta_{x, y}\delta_{x', y'}$ \cite{Shun:eem}.

\subsubsection*{Derivation of the simple dual representation and the variance-based estimation}
Here we derive the simple dual representation $\sigma_{\rm Simple}$. We also show that ${\rm Var}(J_{\bm d})/2{\rm d}t$ gives the entropy production rate when ${\bm d} = {\bm F}$, and reveal its small statistical error as an estimator.\\ \indent
The simple dual representation $\sigma_{\rm Simple}$ can be derived from Eq.~(\ref{eq: dual_ep}) by assuming overdamped Langevin dynamics.
We define $\mathcal{F}''$ as a set of functions $h\in\mathcal{F}$ such that they are written as $h({\bm x}, {\bm x}') = {\bm d}\left(\frac{{\bm x} + {\bm x}'}{2}\right)\;\left({\bm x}' - {\bm x}\right) =: J_{\bm d}$. Then, we derive $\sigma_{\rm Simple}$ as follows:
\begin{subequations}
\begin{eqnarray}
\sigma {\rm d}t &=& \max_{h\in\mathcal{F}'}\left<h - e^{-h} + 1\right>\\
&=& \max_{h\in\mathcal{F}''}\left<h - e^{-h} + 1\right>\label{eq: dual_Langevin1}\\
&=& \max_{{\bm d}}\left[2\left<J_{\bm d}\right> -\frac{{\rm Var}(J_{\bm d})}{2}\right]\label{eq: dual_Langevin2},
\end{eqnarray}
\end{subequations}
where $h^*\in\mathcal{F}''$ is used in Eq.~(\ref{eq: dual_Langevin1}), and $\left<e^{-J_{\bm d}}\right> = 1 - \left<J_{\bm d}\right> + \frac{{\rm Var}(J_{\bm d})}{2} + o({\rm d}t)$ is used in Eq.~(\ref{eq: dual_Langevin2}).\\ \indent
The expansion of $\left<e^{-J_{\bm d}}\right>$ can be proved by using the fact that only the first two cumulants of $J_{\bm d}$ are $O({\rm d}t)$ and the higher order cumulants are $O({\rm d}t^2)$ as shown below.
First, the generalized current under the overdamped Langevin equation defined in Eq.~(\ref{eq: Langevin}) of the main text is written as
\begin{subequations}
\begin{eqnarray}
J_{\bm d} &=& \sum_i d_i({\bm x}(t), t) \circ {\rm d}x_i(t)\\
&=& \sum_i \frac{d_i({\bm x}(t+{\rm d}t), t+{\rm d}t) - d_i({\bm x}(t), t)}{2}{\rm d}x_i(t) + d_i({\bm x}(t), t){\rm d}x_i(t)\\
&=& \frac{1}{2}\sum_{i, j}[\nabla_j d_i({\bm x}(t), t)]{\rm d}x_j(t){\rm d}x_i(t) + \sum_i d_i({\bm x}(t), t){\rm d}x_i(t) + O({\rm d}t^{\frac{3}{2}})\\
&=& \frac{1}{2}\sum_{i, j}[\nabla_j d_i({\bm x}(t), t)]D_{ij}{\rm d}t + \sum_i d_i({\bm x}(t), t)(A_i{\rm d}t + \sum_l \sqrt{2}B_{il}{\rm d}w_l) + O({\rm d}t^{\frac{3}{2}}),
\end{eqnarray}
\end{subequations}
where ${\rm d}{\bm w} := {\bm\eta}(t){\rm d}t$. Then, the means of $J_{\bm d}, J_{\bm d}^2$, and $J_{\bm d}^3$ are evaluated as follows:
\begin{subequations}
\begin{eqnarray}
\left<J_{\bm d}\right> &=& \int {\rm d}{\bm x} p({\bm x}, t)\delta({\bm x}(t) - {\bm x})J_{\bm d}\\
&=& {\rm d}t \int{\rm d}{\bm x}\left\{\sum_{i, j}-d_i({\bm x}, t)\nabla_j\left[D_{ij}p({\bm x}, t)\right] + \sum_i d_i({\bm x}, t)A_i p({\bm x}, t)\right\}\\
&=& {\rm d}t \int{\rm d}{\bm x}{\bm d}({\bm x}, t)^{\mathsf{T}}{\bm j}({\bm x}, t),\label{eq: mean Jd}\\
\left<J_{\bm d}^2\right> &=& \int {\rm d}{\bm x} p({\bm x}, t)\delta({\bm x}(t) - {\bm x})J_{\bm d}^2\\
&=& 2{\rm d}t\int {\rm d}{\bm x} p({\bm x}, t){\bm d}^\mathsf{T}{\bm D}{\bm d},\label{eq: var Jd}\\
\left<J_{\bm d}^3\right> &=& \int {\rm d}{\bm x} p({\bm x}, t)\delta({\bm x}(t) - {\bm x})J_{\bm d}^3\\
&=& {\rm d}t^2 \int {\rm d}{\bm x}p({\bm x}, t)\delta({\bm x}(t) - {\bm x})\left\{\frac{1}{2}\sum_{i, j}[\nabla_j d_i({\bm x}(t), t)]D_{ij} + \sum_i d_i({\bm x}(t), t)A_i\right\}\left\{2{\bm d}^\mathsf{T}{\bm D}{\bm d}\right\},
\end{eqnarray}
\end{subequations}
where only the leading order terms are maintained. 
We note that ${\rm Var}(J_{\bm d}) = \left<J_{\bm d}^2\right>$ holds to order ${\rm d}t$.
We can show that $\left<J_{\bm d}^n\right> = O({\rm d}t^2)$ for any $n \geq 4$ in a similar manner.\\ \indent
Next, we explain the idea of the variance-based estimation. It is straightforward to show that $\left<J_{\bm F}\right> = \sigma{\rm d}t$ and ${\rm Var}(J_{\bm F}) = 2\sigma{\rm d}t$ hold by substituting Eq.~(\ref{eq:Ffield}) into Eq.~(\ref{eq: mean Jd}) and Eq.~(\ref{eq: var Jd}). Thus, the variance of the generalized current can be used as an estimator of the entropy production rate after the training of the model function. The variance-based estimation has an advantage over the variational representations in terms of the statistical error. To evaluate its statistical error as an estimator, we explicitly write the estimated mean and variance of the generalized current as
\begin{eqnarray}
\widehat{\left<J_{\bm d}\right>} &=& \frac{1}{N}\sum_i J_{\bm d}^{(i)},\\
\widehat{{\rm Var}(J_{\bm d})} &=& \frac{1}{N-1}\sum_i \left(J_{\bm d}^{(i)}\right)^2 - \frac{1}{N(N-1)}\left(\sum_i J_{\bm d}^{(i)}\right)^2,
\end{eqnarray}
where $N$ is the number of realizations.
The standard deviation of the estimated mean value satisfies ${\rm std}\left[\widehat{\left<J_{\bm d}\right>} \right] \approx \sqrt{\frac{{\rm d}t}{N}}$ because of ${\rm Var}(J_{\bm d})\propto {\rm d}t$ and the central limit theorem. Similarly, it can be shown that ${\rm std}\left[\widehat{{\rm Var}(J_{\bm d})} \right] \approx \sqrt{\frac{{\rm d}t^2}{N}}$ using the fact that ${\rm Var}(J_{\bm d}^2) = \left<J_{\bm d}^4\right> - \left<J_{\bm d}^2\right>^2 \propto {\rm d}t^2$.
Thus, the variance-based estimation has less statistical error than the variational representations under the short-time condition if we know the thermodynamic force accurately.\\ \indent
Here, the variance-based estimation is only effective for the continuous-time inference with the variational representations $\widehat{\sigma}_{\rm NEEP}$ or $\widehat{\sigma}_{\rm Simple}$ due to the following reasons. First, the estimate of the thermodynamic force ${\bm d}({\bm x} | {\bm\theta}^*)$ already has a statistical error comparable to that of $\widehat{\left<J_{\bm d}\right>}$, which comes from the estimation of a variational representation, for the instantaneous-time inference. On the other hand, for the continuous-time inference, the trained model function ${\bm d}({\bm x}, t | {\bm\theta}^*)$ estimates the thermodynamic force more accurately beyond the single-step current fluctuations, and hence the variance-based estimation can be effective.
Second, the variance-based estimation becomes equivalent to the variational representation $\widehat{\sigma}_{\rm TUR}$ if $\widehat{\sigma}_{\rm TUR}$ is used for training the model function. This is because $\widehat{{\rm Var}(J_{c{\bm d^*}})}/2{\rm d}t$ with the correction term $c = 2\widehat{\left<J_{\bm d^*}\right>}/\widehat{{\rm Var}(J_{\bm d^*})}$ is equivalent to $\widehat{\sigma}_{\rm TUR}$ by the relation $\widehat{{\rm Var}(J_{c{\bm d^*}})}/2{\rm d}t = c^2\widehat{{\rm Var}(J_{\bm d^*})}/2{\rm d}t = 2\widehat{\left<J_{\bm d^*}\right>}^2/{\rm d}t\widehat{{\rm Var}(J_{\bm d^*})}$.

\subsubsection*{Comparison between the variational representations}
\begin{figure}[t]
\includegraphics[width = 0.85\linewidth]{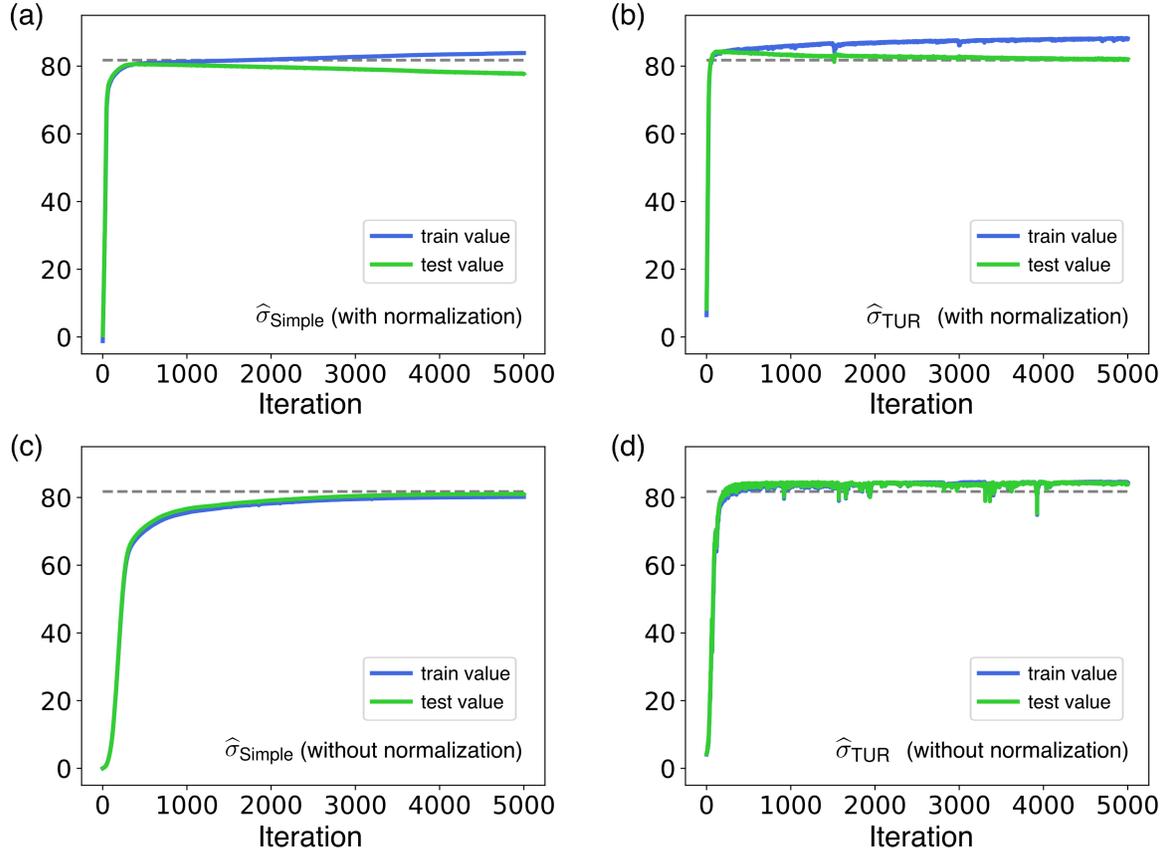}
\caption{\textcolor{black}{Comparison of the learning curves with the following conditions: (a) the simple dual representation (Eq.~(\ref{eq: simple})) with the normalization of data (Eq.~(\ref{eq: normalization})), (b) the TUR (Eq.~(\ref{eq:TURinf})) with the normalization, (c) the simple dual representation without the normalization, and (d) the TUR without the normalization. The blue line is the training value and the green line is the test value, and the dotted line is the true value $\sum_i\sigma(i\Delta t)/M$. The adaptation model is used and the system parameters are set to the same values as those in Fig.~\ref{fig: main2}. The learning curves of the NEEP (Eq.~(\ref{eq: NEEP})) are not shown here, but they are almost the same as those of the simple dual representation. These plots show that the TUR achieves the fastest convergence, while the normalization of data accelerates the convergence in both the cases.}}
\label{fig: supply2}
\end{figure}
\begin{figure}[t]
\includegraphics[width = 0.88\linewidth]{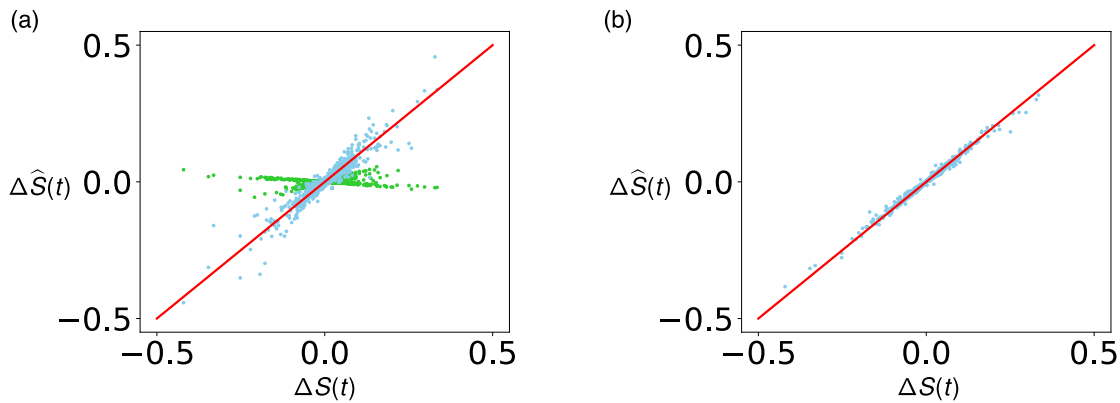}
\caption{Comparison of the entropy production estimations in the breathing parabola model: (a) scatter plot between the true and estimated entropy production using $\widehat{\sigma}_{\rm TUR}$ (Eq.~(\ref{eq:TURinf})), and (b) that using $\widehat{\sigma}_{\rm Simple}$ (Eq.~(\ref{eq: simple})). 
In panel (a), the green dots deviating from the diagonal line are before the correction of $c(t)$, and the blue dots are after the correction. In each plot, the estimated (Eq.~(\ref{eq: dS})) and the true  (Eq.~(\ref{eq: dS_true})) single-step entropy production along 10 trajectories are calculated. The system parameters are set to the same as those in Fig.~\ref{fig: main1}, \textcolor{black}{and the model function is trained using $10^4$ trajectories.} As these plots show, the correction of $c(t)$ in $\widehat{\sigma}_{\rm TUR}$ typically increases the statistical error.
}
\label{fig: supply3}
\end{figure}
Here we compare the performance of the variational representations numerically. \textcolor{black}{We also show that the normalization of trajectory data improves the learning.\\ \indent
In Fig.~\ref{fig: supply2}, we compare the learning curves from the simple dual representation or the TUR with or without the normalization of trajectory data defined in Eq.~(\ref{eq: normalization}). The green line is the test value, and the blue line is the training value which typically takes a larger value than the test value (their expressions are the same:  $\sum_i\widehat{\sigma}(i\Delta t)/M$). The results show that $\widehat{\sigma}_{\rm TUR}$ achieves faster convergence than $\widehat{\sigma}_{\rm Simple}$, which is consistent with the discussion around Eq.~(\ref{eq: comparison}) of the main text.
We also find that the normalization of trajectory data improves the convergence of the learning curve in both the variational representations. The normalization especially reduces the iterations necessary for the test value to converge in the case of the simple dual representation.
We note that the results of the NEEP are not shown here, but they are close to those of the simple dual representation. \\ \indent}
On the other hand, $\widehat{\sigma}_{\rm NEEP}$ and $\widehat{\sigma}_{\rm Simple}$ have an advantage in estimating the thermodynamic force. $\widehat{\sigma}_{\rm TUR}$ requires a correction by the factor $c(t)$ to estimate the thermodynamic force exactly, since the optimal coefficient satisfies ${\bm d}^*({\bm x}, t) = c(t){\bm F}({\bm x}, t)$. The correction by the relation $2\left<J_{\bm d}\right>/{\rm Var}(J_{\bm d})= 1/c(t)$ can decrease the benefit of the continuous-time inference since it is based on the small ensemble  at time $t$.
In Fig.~\ref{fig: supply3}, we compare $\widehat{\sigma}_{\rm TUR}$ and $\widehat{\sigma}_{\rm Simple}$ in terms of the entropy production estimation. We can see that the dots after the correction in (a) scatter more than those in (b), while those before the correction in (a) seem to have similar variance to those in (b).\\ \indent
Finally, we remark on the bias problem of the estimators. The variational representations are biased in the sense that even if ${\bm d}({\bm x}, t) = {\bm F}({\bm x}, t)$ is used, the objective function can be systematically shifted from $\sigma(t)$ when the number of trajectories is small. A criterion to judge the bias would be that the mean of $\widehat{\left<J_{\bm F}\right>}$, which is $\simeq\sigma{\rm d}t$, is comparable to its standard deviation $\simeq \sqrt{\frac{2\sigma{\rm d}t}{N}}$. Here, $N$ is the number of single transitions and ${\rm d}t$ is the time interval used to calculate $\widehat{\left<J_{\bm F}\right>}$ (${\rm d}t$ is used here to distinguish it from the sampling interval $\Delta t$). Thus, when the system is close to equilibrium, we should consider using a larger time step ${\rm d}t$ ($=n\Delta t$) to calculate the generalized current.

\subsection*{Supplementary Note 2: Calculating the analytical solutions}
The analytical solutions for the entropy production rate $\sigma(t)$ as well as for the thermodynamic force ${\bm F}({\bm x},t)$ presented in this paper, are obtained by exactly solving for the instantaneous probability density $p({\bm x},t)$. 
\subsubsection*{Breathing parabola model}
The breathing parabola model described by Eq.\ \eqref{eq:BreathingP} is a colloidal system that remains in a time-dependent non-equilibrium state during the process. For this model, we obtain $p(x,t)$ under a Gaussian ansatz by computing ${\rm Var}({\bm x})$ using a path integral technique \cite{Manikandan:2017awd}. We corroborate this ansatz by checking that this  Gaussian solution does indeed solve the Fokker-Planck equation.   
Note that we could also presumably obtain $p(x,t)$  directly from the Fokker-Planck equation, as we do for the second model in the following section. However, the method we present here is easier in our opinion, for the case when the system remains in a transient state.

In order to compute ${\rm Var}(x)$ for the breathing parabola model, in the path integral formalism, we first write down the moment generating function of $x^2(\tau)$ as a path integral,
\begin{align}
    \left\langle e^{\lambda\; x^2(\tau)}\right\rangle=\int_{x(0)}\int_{x(\tau)} D[x(\cdot)]\;e^{-S[x(\cdot),\lambda]},
\end{align}
where the action $S[x(\cdot), \lambda]$ has the information about the initial conditions of the system, the equations governing the dynamics as well as the quantity $x(\tau)$ whose moment generating function we are interested in computing \cite{Manikandan:2017awd}. For this particular case, the action is given by
\begin{align}
\begin{split}
    S[x(\cdot),\lambda]=V(x(0))+\frac{1}{4T}\int_0^\tau {\rm d}t\;\left(\dot{x}(t)+\kappa(t) x(t)\right)^2 +\lambda x^2(\tau).
    \end{split}
\end{align}
After several partial integration, we can write the action in a manifestly quadratic form as
\begin{align}
    \begin{split}
    S[x(\cdot),\lambda]&=\int_0^\tau {\rm d}t\;x(t) O(t) x(t) + \text{Boundary terms,}
    \end{split}
\end{align}
where the operator $O(t)$ is given by
\begin{align}
    O(t):=-\frac{{\rm d}^2}{{\rm d}t^2}-\dot{\kappa}(t)+\kappa^2(t).
\end{align}
The boundary terms can further be written down as
\begin{align}
\begin{split}
    \text{Boundary terms} = \left(\begin{array}{cc}
         x(0) &\dot{x}(0) \end{array}\right) M\left(\begin{array}{c}
         x(0)\\ \dot{x}(0) \end{array}\right)
  + \left(\begin{array}{cc}
         x(\tau) &\dot{x}(\tau) \end{array}\right) N\left(\begin{array}{c}
         x(\tau)\\ \dot{x}(\tau) \end{array}\right),
         \end{split}
\end{align}
where
\begin{align}
    M= \left(
\begin{array}{cc}
 \frac{\gamma  \kappa (0)}{4 k_B T} & -\frac{\gamma }{4k_B T} \\
 0 & 0 \\
\end{array}
\right),  ~~~~~N= \left(\begin{array}{cc}
         0&0  \\
       \frac{\gamma  \kappa (\tau)}{4k_B T}+\lambda &\frac{\gamma }{4 k_B T} 
    \end{array}\right)
\end{align}
Direct evaluation of the Gaussian integral then gives
\begin{align}
\label{detratio}
        \left\langle e^{\lambda\;x^2(\tau)}\right\rangle=\sqrt{\frac{\det O(t) \vert_{\lambda=0}}{\det O(t)}}.
\end{align}
The determinants appearing in the above expression are \textit{functional determinants}. A method for evaluating the determinant ratio appearing in Eq.\ \eqref{detratio} was introduced in Ref.~\cite{Kirsten:2003fdc}. It was shown that, if $F_\lambda(l)$ is the characteristic polynomial corresponding to the operator $O(t)$, then
\begin{align}
\label{result}
    \sqrt{\frac{\det O(t) \vert_{\lambda=0}}{\det O(t)}}=\sqrt{\frac{F_0(0)}{F_\lambda(0)}}.
\end{align}
There is a natural choice for the identification of the characteristic polynomial $F$, in terms of the matrices $M$ and $N$ as well as the fundamental solutions $z_i(t)$ of the differential operator $O(t)$ such that $O(t) z(t)=0$. In this particular case, the two independent solutions of the equation $O(t) z(t)=0$ are given by
\begin{align}
    z_1(t)=(\alpha  t+1)^2, ~~~~~z_2(t)=\frac{1}{\alpha  t+1}.
\end{align}
The characteristic polynomial can then be obtained as
\begin{eqnarray}
    F_\lambda(0)&=&\det \left[ M+NH(\tau) H^{-1}(0)\right],\\
    H(t)&=&\left(\begin{array}{cc}
         z_1(t)&z_2(t)  \\
         \dot{z}_1(t)& \dot{z}_2(t)
    \end{array}\right).
\end{eqnarray}
\indent
The moment generating function can then be obtained as, 
\begin{align}
  \left\langle e^{\lambda\;x^2(\tau)}\right\rangle=  \sqrt{3} \sqrt{\frac{\alpha  \gamma  (\alpha  \tau+1)^2}{2 k_B T \lambda \left(2 \alpha ^3 \tau^3+6 \alpha ^2 \tau^2+6 \alpha  \tau+3\right)+3 \alpha  \gamma  (\alpha  \tau+1)^2}}
\end{align}
From the moment generating function, it is then straightforward to compute $\langle x^2(\tau)\rangle$. We obtain
\begin{align}
    \langle x^2(\tau)\rangle=\frac{k_BT \left(2 \alpha ^3 \tau^3+6 \alpha ^2 \tau^2+6 \alpha  \tau+3\right)}{3 \left(\alpha  \gamma  (\alpha  \tau+1)^2\right)}.
\end{align}
The instantaneous probability density is therefore given by
\begin{align}
    p(x,\tau) = \frac{e^{-\frac{3 \left(\alpha  \gamma  (\alpha  \tau+1)^2\right)\;x^2}{2k_BT \left(2 \alpha ^3 \tau^3+6 \alpha ^2 \tau^2+6 \alpha  \tau+3\right)}}}{\sqrt{\frac{2 \pi\; k_BT \left(2 \alpha ^3 \tau^3+6 \alpha ^2 \tau^2+6 \alpha  \tau+3\right)}{3 \left(\alpha  \gamma  (\alpha  \tau+1)^2\right)}}}.
\end{align}
It is then straightforward to check that this $p(x,t)$ solves the corresponding Fokker-Planck equation. In order to obtain $\sigma(t)$, we first compute the instantaneous current $j(x,t)$ as
\begin{align}
   j(x,t)= \left( -\kappa(t) x-\frac{k_B T}{\gamma}\partial_x \right)p(x,t).
\end{align}
Then by using Eq.\ \eqref{eq:sigmatime} and Eq.\ \eqref{eq:Ffield} in the main text, we get
\begin{align}
    \sigma(t)&=\int {\rm d}x\;\frac{j(x,t)^2}{D\;p(x,t)} =\frac{\alpha ^3 t^2 \left(\alpha ^2 t^2+3 \alpha  t+3\right)^2}{3 (\alpha  t+1)^4 \left(2 \alpha ^3 t^3+6 \alpha ^2 t^2+6 \alpha  t+3\right)}
\end{align}
which is plotted as the black line in Fig.~\ref{fig: main1}(c). In Fig.~\ref{fig: main1}(b), the black line corresponds to the total entropy production, along a trajectory $\{x_{i\Delta t}\}_{i=0}^N$,
\begin{align}
    S[x(\cdot), t]=\sum_{i=0}^{t/\Delta t-1} F\left(\frac{x_{i\Delta t}+x_{(i+1)\Delta t}}{2}, \left(i + \frac{1}{2}\right)\Delta t\right)\;\left( x_{(i+1)\Delta t}-x_{i\Delta t}\right),
\end{align}
with
\begin{align}
    F(x,t)=\frac{j(x,t)}{D\;p(x,t)}.
\end{align}
\subsubsection*{Biological model}
The second model we have studied is a linear diffusive system of the form (Eq.~(\ref{eq:bioeq}) of the main text): 
\begin{align}
\label{eq:genL}
    \dot{{\bm x}}(t)={\bm A}(t)\;{\bm x}(t)+{\bm B}\cdot {\bm \eta} (t).
\end{align}
It is again the linearity of this system which enables us to solve it exactly. In general dynamical systems of the kind Eq.\ \eqref{eq:genL} are called generalized \"{O}rnstein-Uhlenbeck processes. The corresponding Fokker-Planck equation satisfied by $p({\bm x},t)$ reads
\begin{align}
    \frac{\partial p}{\partial t} = -{\bm \nabla}\left[{\bm A} {\bm x} p({\bm x},t)+ {\bm D}{\bm \nabla}p({\bm x},t)\right],
\end{align}
where $\bm D$ is the diffusion matrix,
\begin{align}
    {\bm D}=\frac{1}{2}{\bm B}{\bm B}^T.
\end{align}
General techniques have been developed in the literature to solve such Fokker-Planck equations \cite{risken1996fokker,Argun:2017erm}, especially for systems reaching a stationary state. Specifying the initial time as $t_0$ and the initial position as $x_0$, the general solution of the Fokker-Planck equation can be obtained as \cite{Argun:2017erm},
\begin{align}
\label{insta}
    p({\bm x},t\vert {\bm x}_0,t_0)=\frac{e^{-\frac{1}{2}\left[ {\bm x}-e^{-(t-t_0){\bm A}}{\bm x}_0\right]^T{\bm S}^{-1}(t-t_0)\left[{\bm x}-e^{-(t-t_0){\bm A}}{\bm x}_0\right]}}{\sqrt{(2 \pi )^2 \det{\bm S}(t-t_0)}},
\end{align}
where the matrix ${\bm S}$ is given by
\begin{align}
    {\bm S}(t)={\bm S}(\infty)-e^{-t{\bm A}}{\bm S}(\infty)e^{t{\bm A}^T},
\end{align}
and the matrix ${\bm S}(\infty)$ can be obtained by solving 
\begin{align}
    {\bm A}{\bm S}(\infty)+{\bm S}(\infty){\bm A}^T = 2{\bm D}.
\end{align}
The instantaneous density $p({\bm x},t)$ can be obtained from Eq.\ \eqref{insta} by integrating out the initial variables ${\bm x}_0$ over the specific initial density. 
In particular, the stationary density if it exists, is given by, 
\begin{align}
\label{eq:pss}
    p_{\rm ss}({\bm x})=\frac{e^{-\frac{1}{2}{\bm x}^T {\bm S}^{-1}(\infty){\bm x}}}{\sqrt{(2 \pi )^2 \det {\bm S}(\infty)}}.
\end{align}
\indent
For the biological model we studied in Eq.\ \eqref{eq:bioeq}, $\beta l(t)=0$ for $t<0$ and $\beta l(t)=0.01$ for $t\geq 0$. For $t<0$,  the equations of motion read
\begin{subequations}
\begin{eqnarray}
\dot{a}(t) &=&-\frac{1}{\tau_a}\left[a(t) - \alpha m(t) \right] + \sqrt{2\Delta_a}\eta_a(t),\\
\dot{m}(t) &=&-\frac{1}{\tau_m}a(t) + \sqrt{2\Delta_m}\eta_m(t).
\end{eqnarray}
\end{subequations}
The above formalism can be straightforwardly applied to compute the corresponding stationary density of this system, which is the initial density at $t=0$.
When $t \geq 0$, the coupled equations read,
\begin{subequations}
\begin{eqnarray}
\dot{a}(t) &=&-\frac{1}{\tau_a}\left[a(t) - \alpha m(t) +0.01\right] + \sqrt{2\Delta_a}\;\eta_a(t),\\
\dot{m}(t) &=&-\frac{1}{\tau_m}a(t) +\sqrt{2\Delta_m}\; \eta_m(t).
\end{eqnarray}
\end{subequations}
To apply the above formalism to $t>0$, we make the change of variables $\alpha m -0.01 = m^\prime$. In the new variables, the equations read
\begin{subequations}
\begin{eqnarray}
\dot{a}(t) &=&-\frac{1}{\tau_a}\left[a(t) - m^\prime(t) \right] + \sqrt{2\Delta_a}\eta_a(t),\\
\dot{m^\prime}(t) &=&-\frac{\alpha}{\tau_m}a(t) + \alpha \sqrt{2\Delta_m}\;\eta_m(t).
\end{eqnarray}
\end{subequations}
In this form, we can now apply the formalism described above to obtain the instantaneous density in Eq.\ \eqref{insta} in the transformed variables  ${a,\;m^{\prime}}$. We then need to revert back to the original variables ${a,\;m}$ using $m=\frac{m^\prime+0.01}{\alpha}$ and integrate out  the initial stationary density obtained for the $t<0$ case to obtain $p(a,m,t)$. Since the intermediate densities are all Gaussian, we take care of the Jacobian factors under coordinate transformation by making sure that the densities are properly normalized. 

Once we have $p(a,m,t)$, we obtain ${\bm j}(a,m,t)$, ${\bm F}(a,m,t)$, and $\sigma(t)$, using Eqs.~(\ref{eq:prob_cur}) - (\ref{eq:Ffield}) in the main text. The expression are then used to plot the analytical results in Fig.~\ref{fig: main2}.

\subsubsection*{Parameters for the breathing parabola model}
\textcolor{black}{For the breathing parabola model, in Fig.~\ref{fig: main1}, we used the following parameter values, adapted from \cite{Blickle:2012rms}. 
\begin{align}
    \begin{split}
       \text{Boltzmann constant, } k_B &= 1.38 \times 10^{-23}~{\rm J}{\rm K}^{-1},\\
        \text{Temperature of the medium (water),  }T &= 298~{\rm K}\\
       \text{Viscosity of water, } \eta &= 8.9 \times 10^{-4}~{\rm kg m^{-1}s^{-1}}\\
        \text{Radius of the trapped particle, } r &= 1.45 \times 10^{-6}~{\rm m}\\
        \text{Viscous drag, } \gamma &= 6\pi \eta r\\
        \text{Diffusion constant, } D&= \frac{k_B T }{\gamma}\\
        \text{Trap stiffness at $t = 0$, } \kappa (t = 0) &= 281~\text{fN}\;\mu {\rm m}^{-1}.
    \end{split}
\end{align}
The particular protocol we chose is $\kappa (t) =\gamma\alpha/(1+\alpha t)$. We chose the parameter $\alpha =11.5358~{\rm s}^{-1}$, so that at $t = 0$, $\kappa(0) =281~{\rm fN\;\mu}{\rm m}^{-1}$. }
\end{widetext}

\end{document}